\documentclass{jltp}

\def\beq{\begin{equation}}
\def\eeq{\end{equation}}
\def\bleq{\begin{eqnarray}}
\def\eleq{\end{eqnarray}} 
\def\bfig{\begin{figure}}
\def\efig{\end{figure}}
\def\bline{\begin{multline}}
\def\eline{\end{multline}}

\newcommand{\mean}[1]{\langle #1 \rangle}

\def\eps{\epsilon}

\def\half{\frac{1}{2}}

\def\k{{\bf k}}

\def\q{{\bf q}}

\def\O{{\bf O}} 
 
\def\Q{{\bf Q}} 

\def\S{{\bf S}}

\def\para{\parallel}

\def\kperp{{k_\perp}} 
\def\kperpp{{k_\perp'}} 
\def\qperp{{q_\perp}}

\def\w{\omega}

\usepackage{graphicx}
\usepackage{amssymb}
\usepackage{amsmath}           

\graphicspath{{figures/}}

\title{Superconductivity and Antiferromagnetism in Quasi-one-dimensional
  Organic Conductors}

\author{N. Dupuis$^{1,2}$, C. Bourbonnais$^3$ and J. C. Nickel$^{2,3}$} 

\address{$^1$ Department of Mathematics, Imperial College, \\
180 Queen's Gate, London SW7 2AZ, United Kingdom \\
$^2$ Laboratoire de Physique des Solides, CNRS UMR 8502, \\
 Universit\'e Paris-Sud, 91405 Orsay, France \\
$^3$ Regroupement Qu\'ebecois sur les Mat\'eriaux de Pointe, \\
Universit\'e de Sherbrooke, Sherbrooke,
  Qu\'ebec, Canada J1K-2R1}

\runninghead{N. Dupuis, C. Bourbonnais and J. C. Nickel}{Superconductivity and
  Antiferromagnetism in Organic Conductors}

\begin{document}

\maketitle

\begin{abstract}
We review the current understanding of superconductivity in the
quasi-one-dimensional organic conductors of the Bechgaard and Fabre salt
families. We discuss the interplay between
superconductivity, antiferromagnetism, and  
charge-density-wave fluctuations. The  connection to recent experimental
observations supporting unconventional pairing and the possibility of a
triplet-spin order parameter for the superconducting phase is also presented.
  
\end{abstract}

\section{INTRODUCTION}

Superconductivity in organic conductors was first discovered in the ion
radical salt (TMTSF)$_2$PF$_6$.\cite{Jerome80} Later on, it 
was found in most Bechgaard [(TMTSF)$_2$X] and Fabre [(TMTTF)$_2$X]
salts. These  salts are based on the organic molecules 
tetramethyltetraselenafulvalene (TMTSF) and tetramethyltetrathiafulvalene
(TMTTF).  The monovalent anion X can be either a centrosymmetric (PF$_6$,
AsF$_6$, etc.) or a non-centrosymmetric (ClO$_4$, ReO$_4$, NO$_3$, FSO$_3$,
SCN, etc.) 
inorganic molecule. (See Refs.~\onlinecite{Jerome82,Ishiguro90} for previous
reviews on these compounds.) Although they are definitely not ``high-$T_c$''
superconductors -- the 
transition temperature is of the order of 1 K --, these quasi-one-dimensional
(quasi-1D) conductors share several properties of high-$T_c$ superconductors
and other strongly-correlated electron systems such as layered organic
superconductors\cite{McKenzie98,Lefebvre00} or heavy-fermion
materials.\cite{Flouquet05}  The metallic phase of all these conductors
exhibits unusual 
properties which cannot be explained within the framework of Landau's Fermi
liquid theory and remain to a large extent to be understood. The
superconducting phase is unconventional (not $s$-wave). Magnetism is
ubiquitous in these correlated systems and might provide the key to the
understanding of their behavior. 

The quest for superconductivity in organic conductors was originally motivated
by Little's proposal that highly polarizable molecules could lead -- {\it via}
an excitonic pairing mechanism -- to tremendously large transition
temperatures.  Early efforts towards the chemical synthesis of such compounds
were not successful -- as far as superconductivity is concerned --, but lead
to the realization of a 1D charge transfer salt (TTF-TCNQ) undergoing a Peierls
instability at low temperatures.\cite{Jerome04} Attempts to suppress the
Peierls state and stabilize a conducting (and possibly superconducting) state
by increasing the 3D character of this 1D conductor proved to be unsuccessful. 

Organic superconductivity was eventually discovered in
the Bechgaard salt (TMTSF)$_2$PF$_6$ under 9~kbar of pressure.\cite{Jerome80}
It was subsequently found in  other members of the  (TMTSF)$_2$X series. 
Most of the Bechgaard salts are insulating at ambient pressure and low
temperatures,\cite{Bechgaard80} and it 
came as a surprise that the insulating state of these materials is a
spin-density-wave (SDW) rather than an ordinary Peierls state.\cite{Jerome82}
The important part played by magnetism in these compounds was further
revealed when it was found that their  phase diagram  only shows   a part of a
larger sequence of ordered states, which   includes the N\'eel and  the
spin-Peierls phases  of   their sulfur analogs, the Fabre salts (TMTTF)$_2$X
series.\cite{Bourbonnais99}   

\begin{figure}
\centerline{\includegraphics[bb=95 570 400 695,width=12cm]{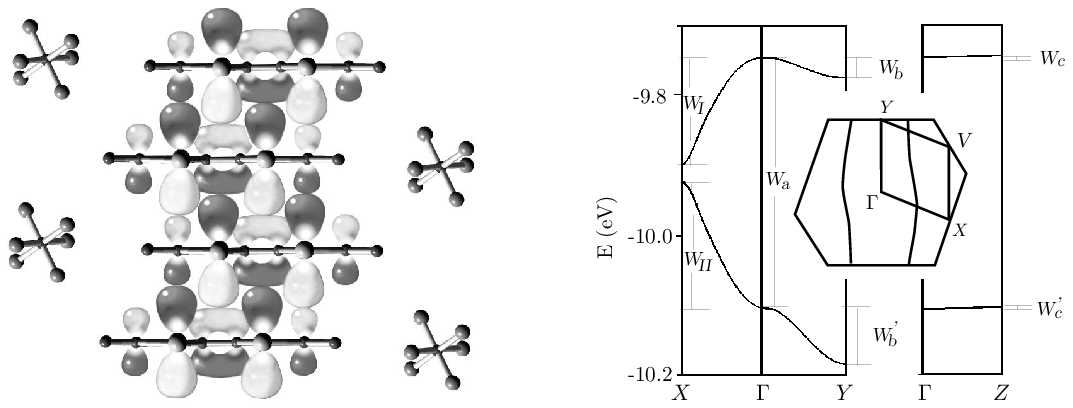}}
\caption{(Left) A side view of the Bechgaard/Fabre salt crystal structure with
  the electron orbitals of the organic stacks (courtesy of
  J. Ch. Ricquier). (Right) Electronic dispersion 
  relation and projected 2D Fermi surface of (TMTTF)$_2$Br (reprinted with
  permission from Ref.~\onlinecite{Balicas94}. Copyright 1994 by EDP
  Sciences).}  
\label{fig:structure}
\end{figure}

The charge transfer from the organic molecules to the
anions leads to a commensurate band filling 3/4 coming from the 2:1
stoichiometry. The metallic character of these compounds at high enough
temperature is due to the 
delocalization of carriers via the overlap of $\pi$-orbitals between
neighboring molecules along the stacking direction ($a$ axis)
(Fig.~\ref{fig:structure}).\cite{Jerome82} The electronic dispersion
relation obtained from quantum chemistry calculations (extended H\"uckel
method) is well approximated by the following tight-binding
form\cite{Grant83,Yamaji82,Ducasse86,Balicas94} 
\bleq
\eps(\k) &=& -2t_a \cos(k_aa/2)-2t_{\perp b} \cos(k_b b) -2t_{\perp c} \cos(k_c
c) \nonumber \\ 
&\simeq& v_F(|k_a|-k_F) -2t_{\perp b} \cos(k_b b) - 2t'_{\perp b} \cos(k_b b) 
\nonumber \\ && -2t_{\perp c} \cos(k_c c) + \mu , 
\label{dispersion}
\eleq
where it is assumed that the underlying lattice is orthorhombic. This
expression is a simplification of the dispersion relation -- the actual
crystal lattice symmetry is triclinic -- but it retains the essential
features. The conduction band along the chain direction has an overall width
$4t_a$ ranging between 0.4 and 1.2 eV, depending on the organic molecule
(TMTSF or TMTTF) and the anion. As the electronic overlaps in the
  transverse  $b$ and $c$ directions are much weaker than along the organic
stacks, the dispersion law is 
strongly anisotropic, $t_{\perp b}/t_a\simeq 0.1$  and $t_{\perp c}/t_{\perp
  b}\simeq 0.03$, and the Fermi surface consists of two open warped sheets
(Fig.~\ref{fig:structure}). In the second line of Eq.~(\ref{dispersion}), the
electronic dispersion is linearized around the two 1D Fermi points $\pm k_F$,
with $v_F$ the Fermi velocity along the chains ($\mu$ is the chemical
potential). The next-nearest-chain hopping $t'_{\perp b} \propto t_{\perp
  b}^2/t_a + \ldots$ is introduced in order to keep the
shape of the Fermi surface unchanged despite the linearization. 
The anions located in centrosymmetric cavities lie slightly above or below
the molecular planes. This structure leads to a dimerization of the organic
stacks and a (weak) gap $\Delta_D$, thus making the hole-like band effectively
half-filled at sufficiently low energy or temperature.\cite{Emery82,Barisic81}
(See Refs.~\onlinecite{Jerome82,Jerome04,Bourbonnais99} for a detailed
discussion of the structural properties of quasi-1D organic conductors.) In
the presence of interactions, commensurate band-filling introduces Umklapp
scattering, which affects the nature of the possible phases in these
materials.   

\begin{figure}
\centerline{\includegraphics[bb=64 196 569 668,width=8cm]{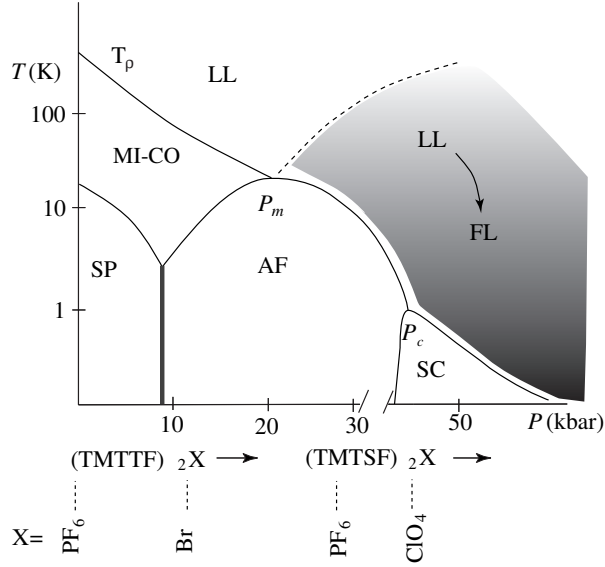}} 
\caption{The generic phase diagram of the Bechgaard/Fabre salts
  as a function of pressure or anion X substitution.
  LL: Luttinger liquid, MI: Mott insulator, CO: Charge order,
  SP: Spin-Peierls, AF: antiferromagnetism, SC: superconductivity, FL: Fermi
  liquid.} 
\label{fig:dia_gene}
\end{figure}

What is remarkable about these electronic systems is the variety of ground
states that 
can be achieved either by chemical means, namely substituting selenium by
sulfur in the organic molecule or changing the nature of the anion (its 
size or  symmetry), or applying pressure (Fig.~\ref{fig:dia_gene}). At low
pressure, members of the sulfur series are Mott insulators (MI) from which
either a lattice distorted spin-Peierls (SP) state --  often  preceded by a
charge ordered (CO) state --  or a commensurate-localized antiferromagnetic
state (AF) can develop. On the other hand, itinerant antiferromagnetism
(spin-density wave (SDW)) or superconductivity is found in the selenide
series. Under pressure, the properties of the Fabre salts evolve towards those
of the Bechgaard salts. The compound (TMTTF)$_2$PF$_6$ spans the entire phase
diagram as pressure increases up to 50 kbar or so
(Fig.~\ref{fig:dia_pf6}),\cite{Jaccard01,Wilhelm01,Adachi00} thus showing the
universality of the phase diagram in
Fig.~\ref{fig:dia_gene}.\cite{Bourbonnais98}  

A large number of both theoretical and experimental works have been devoted to
the understanding of the normal phase and the mechanisms leading to long-range
order at low temperature. The presence of antiferromagnetism over a large
pressure range does indicate that repulsive interactions among carriers are
important. The low-dimensionality of the system is also expected to play a
crucial role. On the one hand, in the presence of repulsive interactions a
strongly anisotropic Fermi surface with good nesting properties is
predominantly unstable against the formation of an SDW state which is
reinforced at low temperature by commensurate band filling. On the other hand,
when the temperature exceeds the transverse dispersion $\sim t_{\perp b}$, 3D
(or 2D) coherent electronic motion is suppressed and the conductor behaves as
if it were 1D; the Fermi
liquid picture breaks down and the system becomes a Luttinger
liquid.\cite{Haldane81,Giamarchi_book} The relevance of 1D physics for the
low-temperature properties ($T\lesssim t_{\perp b}$), as well as a detailed
description of the 
crossover from the Luttinger liquid to the Fermi liquid, is one of the most
important issues in the debate surrounding the theoretical description of the
normal state of these materials. As far as low-temperature phases are
concerned, a chief objective is to reach a good description of the
superconducting phase -- the symmetry of the order parameter is still under
debate -- and the mechanisms leading to superconductivity. Owing to the close
proximity of superconductivity and magnetism in the phase diagram of
Fig.\ref{fig:dia_gene}, it is  essential to first  discuss  the origin of
antiferromagnetism in both series of compounds.

\section{N\'EEL ANTIFERROMAGNETISM AND SPIN-DENSITY WAVE} 

\subsection{Fabre salts at ambient pressure: Mott-insulator regime} 

The Fabre salts (TMTTF)$_2$X  at ambient pressure are located on the left of
the 
phase diagram in Figure \ref{fig:dia_gene}. Both the nature of correlations and
the mechanism of long-range order at low temperature are now rather well
understood. Below the temperature $T_\rho\sim 100$ K (see
Fig.~\ref{fig:dia_gene}), the resistivity develops a thermally activated
behavior\cite{Coulon82a} and the system enters a Mott-insulator regime. The
corresponding charge gap $\Delta_\rho\sim \pi T_\rho$ can be deduced from
$T_\rho$ and turns out to be larger than the (bare) transverse bandwidth
$t_{\perp b}$, which in turn suppresses any possibility of transverse single
particle band motion and makes the system essentially one-dimensional. The
charge gap $2\Delta_\rho$ is also directly observed in the optical 
conductivity.\cite{Schwartz98} The members of the (TMTTF)$_2$X series
thus behave as typical 1D Mott insulators  below $T_\rho$ with the carriers
confined along the organic stacks -- as a result of the Umklapp scattering due
to the commensurability of the electronic density with the underlying
lattice.\cite{Emery82,Barisic81}  
This interpretation agrees with the absence of anomaly in the spin
susceptibility at $T_\rho$,\cite{Wzietek93} in accordance with the spin-charge
separation  characteristic of 1D systems.\cite{Giamarchi_book} It is further
confirmed by measurements of the spin-lattice relaxation rate $1/T_1$. The
Luttinger liquid theory predicts\cite{Bourbonnais89,Bourbonnais93}  
\beq
\frac{1}{T_1} = C_0 T \chi_s^2(T) + C_1 T^{K_\rho} ,
\label{T1}
\eeq
where $C_0$ and $C_1$ are temperature independent constants. $\chi_s(T)$ is
the uniform susceptibility and $K_\rho$ the Luttinger liquid charge stiffness
parameter. The two contributions in (\ref{T1}) correspond to
paramagnons or spinons ($q\simeq 0$) and AF spin fluctuations ($q\simeq
2k_F$). Both the temperature dependence of $\chi_s(T)$ and the presence of
AF fluctuations lead to an enhancement of $1/T_1$ with respect
to the Korringa law  $(T_1T)^{-1}={\rm const}$ which holds in
higher-dimensional metals. In a 1D Mott insulator $K_\rho=0$, which leads to
$T_1^{-1}=C_0 T \chi_s^2(T)+C_1$ in good agreement with experimental
measurements of $T_1$ and $\chi_s$.\cite{Wzietek93} 

The low-energy excitations in the Mott-insulator regime are 1D spin
fluctuations. By lowering the temperature, these fluctuations can 
propagate in the transverse direction and eventually drive an
AF transition. This transition is not connected to Fermi
surface effects. The condition $\Delta_\rho>t_{\perp b}$ precludes a
single-particle coherent motion in the transverse direction, and the concept
of Fermi surface remains ill defined in the Fabre 
salts at ambient pressure. AF long-range order comes from
interchain transfer of bound electron-hole pairs leading to a kinetic exchange
interaction $J_\perp$ between spin densities on neighboring chains -- much in
analogy with the exchange interaction between localized spins in the
Heisenberg limit. An effective Hamiltonian can be derived from a
renormalization group (RG) calculation,\cite{Bourbonnais88,Bourbonnais91}
\beq
H_\perp = J_\perp \int dx \sum_{\mean{i,j}} \S_i(x) \cdot \S_j(x) ,
\;\;\;\;\; 
J_\perp \simeq \frac{\xi_\rho}{a} \frac{t_{\perp b}^{*2}}{\Delta_\rho} ,
\label{Jperp}
\eeq 
where $t_{\perp b}^*$ is the effective interchain hopping at the energy scale
$\Delta_\rho$ and $a$ the lattice spacing along the chain. The sum in
Eq.~(\ref{Jperp}) is over nearest-neighbor chains. The naive value
$t_{\perp b}^{*2}/\Delta_\rho$ of the exchange interaction $J_\perp$ is
enhanced by the factor 
$\xi_\rho/a$ where $\xi_\rho=v_F/\Delta_\rho$ is the intrachain coherence
length induced by the Mott gap along which virtual interchain hoppings can
take place. Within a mean-field treatment of $H_\perp$, the condition for the
onset of long-range order is given by $J_\perp \chi_{\rm 1D}(2k_F,T)=1$ where
$\chi_{\rm 1D}(2k_F,T)\sim (T/\Delta_\rho)^{-1}$ is the exact power law form
of the 1D AF spin susceptibility. This yields a N\'eel temperature
\beq
T_N \sim \frac{t_{\perp b}^{*2}}{\Delta_\rho} . 
\label{TN}
\eeq
Since $T_\rho$ and $\Delta_\rho$ decrease under pressure
(Fig.\ref{fig:dia_gene}), Eq.~(\ref{TN}) predicts an 
increase of $T_N$ with pressure -- assuming a weak pressure dependence of
$t^*_b$ -- as observed experimentally (see Fig.~\ref{fig:dia_gene}). The
relation $T_NT_\rho \sim t_{\perp b}^{*2} \sim {\rm const}$ has been observed
in (TMTTF)$_2$Br.\cite{Brown97}  

\subsection{Bechgaard salts: itinerant magnetism}
\label{subsec:itinerant}

With increasing pressure, $T_\rho$ drops and finally merges with the AF
transition line at $P_m$, beyond which there is no sign of a Mott gap in the
normal phase. 
The Fabre salts then tend to behave similarly to the Bechgaard salts
(Fig.~\ref{fig:dia_gene}). The change of behavior at $P_m$ is usually 
attributed to a deconfinement of carriers, i.e. a crossover from a Mott
insulator to a -- metallic -- Luttinger liquid. At lower temperature,
single-particle transverse hopping is expected to become relevant and induce a
dimensional crossover at a temperature $T_x$ from the Luttinger liquid to
a 2D or 3D metallic state. With increasing pressure, the AF transition
becomes predominantly driven by the instability of the whole warped Fermi
surface due to the nesting mechanism.  
Although there is a general agreement on this scenario, there is considerable
debate on how the dimensional crossover takes place and the nature of the
low-temperature metallic state. 

On the theoretical side, simple RG arguments indicate that the crossover from
the Luttinger liquid to the 2D regime takes place at the
temperature\cite{Bourbonnais84} 
\beq 
T_x \sim \frac{t_{\perp b}}{\pi} \left(
\frac{t_{\perp b}}{t_a}\right)^{\frac{1-K_\rho}{K_\rho}} , 
\label{Tx}
\eeq
where $K_\rho$ is the Luttinger liquid parameter. For non-interacting
electrons ($K_\rho=1$), Eq.~(\ref{Tx}) would give $T_x\sim t_{\perp b}$: the
2D Fermi 
surface is irrelevant when temperature is larger than the dispersion in the
$b$ direction. For interacting electrons ($K_\rho\neq 1$), the interchain
hopping amplitude $t_{\perp b}$ is reduced to an effective value $t_{\perp
  b}^*$ and the dimensional crossover occurs at a lower temperature $T_x\sim
t_{\perp b}^*<t_{\perp b}$. A 
detailed theoretical picture of the dimensional crossover is still lacking. In
particular, whether it is a sharp crossover or rather extends over a wide
temperature range -- as shown by the shaded area in Fig.~\ref{fig:dia_gene} --
is still an open issue.

\subsubsection{The strong-correlation picture}

Some experiments seem to indicate that correlations still play an important
role even in the low-temperature phase of the Bechgaard salts. For instance, a
significant enhancement of $1/T_1T$ with respect to the Korringa law --
although weaker than in the Fabre salts at ambient pressure -- is still
present.\cite{Wzietek93} This behavior has been explained in terms of 1D
spin fluctuations persisting down to the dimensional crossover temperature
$T_x\sim 10$ K, below which the Korringa law is
recovered.\cite{Wzietek93,Bourbonnais93}

The restoration of a plasma edge in the transverse $b'$ direction at low
temperature in (TMTSF)$_2$PF$_6$ -- absent in the Fabre salts -- suggests the
gradual emergence of a coherent motion in the $(ab)$ planes below $T_x \sim
100$ K.\cite{Jacobsen81,Jacobsen83} (${\bf b}'$ is normal to ${\bf a}$ and
${\bf c}$ in the $(ab)$ plane. It differs from ${\bf b}$ due to the triclinic
structure.) However, the frequency dependence of the
optical conductivity is inconsistent with a Drude-like metallic
state.\cite{Cao96,Dressel96,Schwartz98} The low-energy peak carries only 
1\% of the total spectral weight and is too narrow to be interpreted as a
Drude peak with a frequency-independent scattering time. It has been
proposed that this peak is due to a collective mode that bears some
similarities with the sliding of a charge-density wave -- an interpretation
supported by the new phonon features that emerge at low
temperature.\cite{Cao96} 
Furthermore, 99\% of the total spectral weight is found in a finite
energy peak around 200 cm$^{-1}$. It has been suggested that this peak is 
a remnant of a ${1\over 4}$-filled  Mott gap $\Delta_\rho$, observed in the
less metallic Fabre salts at ambient pressure.\cite{Giamarchi97,Schwartz98} In
this picture, (TMTSF)$_2$PF$_6$ 
is close to the border between a Mott insulator and a 
Luttinger liquid, and the low-temperature metallic behavior is made possible 
by the interchain coupling.\cite{Schwartz98,Vescoli98,Giamarchi04} A
different interpretation has been proposed for  the far infrared spectrum in
optical conductivity and is based on the weak half-filling character of the
band for interactions in  the Hubbard limit.\cite{Favand96}  

The longitudinal resistivity in (TMTSF)$_2$PF$_6$ is found to be metallic, 
with a $T^2$ law between the SDW transition and 150 K, crossing over to
a sublinear temperature dependence above 150 K with an exponent in the range
$0.5-1$.\cite{Moser00,Auban99} 
While this observation would be consistent with a
dimensional crossover to a low-temperature Fermi liquid regime taking
place at $T_x\sim 150$ K, the transverse resistance $\rho_b$ along the $b$ axis
apparently fails to show the expected $T^2$ behavior. Given the difficulties
of a direct dc measurement, owing to non-uniform current distributions between
contacts, conflicting results have been published in the
literature.\cite{Mihaly00,Moser00} Nevertheless, below $T\sim 80$ K 
$\rho_b$ can be deduced from $\rho_a\sim T^2$ and $\rho_c\sim T^{1.5}$ using a
tunneling argument, which yields $\rho_c=(\rho_a\rho_b)^{1/2}$ and therefore
$\rho_b\sim T$. Moreover, contactless  -- microwave -- transverse conductivity
measurements in the (TMTSF)$_2$PF$_6$ salt fail to reveal the emergence of a
Fermi liquid $T^2$ temperature dependence of the resistivity in the $b$
direction in this temperature range.\cite{Fertey99}   
As far as $\rho_c$ is concerned, a maximum around $T_{\rm max}\sim 80$ K
has been observed, with a metallic -- though incoherent -- behavior
$\rho_c\sim T^{1.5}$ at lower temperature.\cite{Moser98} $T_{\rm max}$ is
highly sensitive to pressure, whereas the interchain hopping $t_{\perp b}$ is
not. Therefore, $T_{\rm max}$ cannot be directly identified with $t_{\perp
  b}$, but could be related to a -- weakly --
renormalized value $t_{\perp b}^*\sim T_x$ in agreement with predictions of the
Luttinger liquid theory [see Eq.~(\ref{Tx})]. The transport
measurements seem to be indicative of a gradual crossover between a
Luttinger liquid and a Fermi liquid occurring in the temperature range $10-80$
K. The onset of 3D coherence and Fermi liquid behavior would then be
related to the interplane coupling $t_{\perp c}$ between $(a,b)$
planes.\cite{Moser98} 

The absence of Fermi liquid behavior down to very low temperatures in the
Bechgaard salts seems to be further supported by photoemission experiments. 
ARPES fails to detect quasi-particle features or the trace of a Fermi surface
at 150 K.\cite{Zwick97} Similar conclusions were deduced from integrated
photoemission at 50 K.\cite{Dardel93}
However, photoemission results -- e.g. the absence of dispersing structure and
a power-law frequency dependence which is spread over a large energy scale of
the order of 1 eV -- do not conform with the predictions of the
Luttinger theory and might be strongly influenced by surface effects. 

The existence of strong correlations suggests that the kinetic interchain
exchange $J_\perp$, which drives the AF transition in the sulfur series, still
plays an important role in the Bechgaard salts. In this picture,
the decrease of $T_N$ with increasing pressure is due both to the decrease of
$J_\perp$ and the deterioration of the Fermi surface nesting. This scenario is
supported by RG calculations.\cite{Bourbonnais91}  

All the experiments mentioned so far favor different -- and sometimes
incompatible -- scenarios for the dimensional crossover. However, the
high-temperature phase of the Bechgaard salts is always
analyzed on the basis of the Luttinger liquid theory. A consistent
interpretation of the experimental results therefore requires to find a
common $K_\rho$ parameter and to determine the value of the remnant of the Mott
gap $\Delta_\rho$. NMR,\cite{Wzietek93} dc transport,\cite{Moser98,Georges00}
and optical measurements\cite{Schwartz98,Vescoli98} have been interpreted in
terms of the Luttinger theory with $K_\rho\simeq 0.23$ and quarter-filled
Umklapp scattering.\cite{Jerome04,Giamarchi04} This interpretation, as well as
the mere existence of strong correlations, is not without raising a number of
unanswered questions (see the next section). For instance, $K_\rho\simeq
0.23$ would lead according to (\ref{Tx}) to $T_x\sim 10^{-3}t_{\perp b}$, a
value much below the experimental observations.

\subsubsection{The weak-correlation picture}

On the other hand, there are experiments pointing to the absence of strong
correlations in the Bechgaard salts. One of the most convincing arguments 
comes from the so-called Danner-Chaikin oscillations.\cite{Danner94}
Resistance measurements of (TMTSF)$_2$ClO$_4$ in the $c$ 
direction show pronounced resonances when an applied magnetic field is rotated
in the $(ac)$ plane at low temperature. The complete angular dependence of
the magneto-resistance can be reproduced within a semiclassical approach. The
position of the resonance peaks is given by the zeros of the Bessel function
$J_0(\gamma)$ evaluated at $\gamma=2t_{\perp b}cB_x/v_FB_z$ ($c$ is the
interchain spacing in the $c$ direction). This enables a direct measure of the
interchain hopping amplitude in the $b$ direction, yielding $t_{\perp b}\simeq
280$ K above the anion 
ordering transition taking place at 24 K, in very good agreement with values
derived from band calculations.\cite{Grant83,Yamaji82,Ducasse86}   
These results can hardly be reconciled with the existence of strong
correlations. Sizeable 1D fluctuations should lead to a strong ($k_\para,\w$)
dependence of the self-energy, and in turn to a significant renormalization of
$k_\perp$-dependent quantities like the interchain hopping
amplitudes.\cite{Bourbonnais91} This lends support to the idea that the
low-temperature phase of the Bechgaard salts can be described as a weakly
interacting Fermi liquid subject to spin fluctuations induced by the nesting
of the Fermi surface.\cite{Gorkov96,Zheleznyak99} 

The weak-coupling approach has been particularly successful in the framework 
of the Quantized Nesting Model.\cite{Chaikin96,Lederer96,Yakovenko96} The
latter explains the cascade of SDW phases induced by a magnetic field in
(TMTSF)$_2$PF$_6$ and (TMTSF)$_2$ClO$_4$, and provides a natural explanation
for the quantization of the Hall effect -- $\sigma_{xy}=2Ne^2/h$ ($N$ integer)
per $(ab)$ plane -- observed in these phases. Furthermore, it reproduces the
experimental phase diagram only for interchain hopping amplitudes $t_{\perp
b},t_{\perp c}$ close to their unrenormalized values.  

Despite the apparent success of the weak-coupling approach, it has
nevertheless become
clear that the SDW phase of the Bechgaard salts is not conventional. Recent
experiments have shown that the $2k_F$ SDW coexists with a $2k_F$ and a --
weaker -- $4k_F$ charge-density wave (CDW) in
(TMTSF)$_2$PF$_6$.\cite{Pouget96,Kagoshima99} Since there is no $2k_F$ phonon
softening associated to this transition, the emergence of this CDW state
differs from what is  usually seen for an ordinary Peierls state. This
unusual ground-state can be explained on the basis of a quarter-filled 1D
model with dimerization and onsite, nearest-neighbor and next-nearest-neighbor
Coulomb interactions,\cite{Seo97,Kobayashi98,Mazumdar99,Tomio00,Tomio01} but
this explanation remains to be confirmed. 

\subsubsection{The normal phase above the superconducting phase}
\label{subsubsec:np}

\begin{figure}
\centerline{\includegraphics[bb=0 10 450 410,width=7cm]{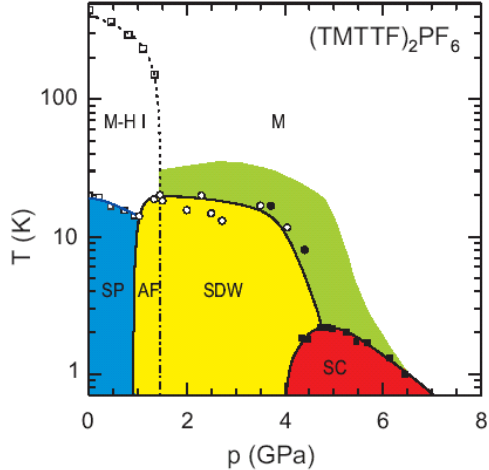}}
\caption{(color online) $(P,T)$ phase diagram of (TMTTF)$_2$PF$_6$. The
  (green) shaded 
  area above the SDW and SC phase indicates the region of the normal phase
  where spin fluctuations are significant. (Reprinted with permission from 
  Ref.~\onlinecite{Wilhelm01}. Copyright 2001 by EDP Sciences.) }
\label{fig:dia_pf6} 
\end{figure}

It is remarkable that the superconducting phase lies next to the SDW
phase -- which is actually a mixture SDW-CDW -- and reaches its maximum
transition temperature $T_c\sim 1$ K at the pressure $P_c$ where $T_{\rm
SDW}$ and $T_c$ join (see Figs.~\ref{fig:dia_gene} and \ref{fig:dia_pf6}). 
In the normal phase above the SDW phase, the resistivity along the $a$ axis
decreases with temperature, reaches a minimum at $T_{\rm min}$, and then shows
an upturn and a strong enhancement related to the proximity of the SDW phase
transition that occurs at $T_{\rm SDW}<T_{\rm min}$. The region of the normal
phase where strong AF fluctuations are present ($T_{\rm SDW}<T<T_{\rm min}$)
extends over the pressure range 
where the ground state is superconducting (Fig.~\ref{fig:dia_pf6}). Its width
in temperature decreases with increasing pressure, so that the
superconducting transition temperature appears to be closely linked to $T_{\rm
  min}$. These observations strongly suggest an intimate relationship between
spin fluctuations and superconductivity in the Bechgaard/Fabre
salts.\cite{Jaccard01,Wilhelm01}  The importance of spin fluctuations above
the superconducting phase is further confirmed by the persistence of the
enhancement of the spin-lattice relaxation rate $1/T_1$ for
$P>P_c$.\cite{Wzietek93} Besides the presence of spin fluctuations at low
temperature, charge fluctuations have also been observed in the normal phase
{\it via} optical conductivity measurements.\cite{Cao96} 

\section{SUPERCONDUCTIVITY}

Some of the early experiments in the Bechgaard salts were not in contradiction
with a conventional BCS superconducting state. For instance, the specific heat
in (TMTSF)$_2$ClO$_4$ obeys the standard temperature dependence $C/T=\gamma +
\beta T^2$ 
above the superconducting transition, and the jump at the transition $\Delta
C/\gamma T_c\simeq 1.67$ is close to the BCS value 1.43. The ratio
$2\Delta(T=0)/T_c\simeq 3.33$, obtained from the gap deduced from the
thermodynamical critical field, is also in reasonable agreement with the
prediction of the BCS theory ($2\Delta/T_c\simeq
3.52$).\cite{Garoche82,Garoche82a} 
Early measurements of $H_{c2}(T)$, performed in the vicinity of the zero-field
transition temperature, were also interpreted on the basis of the BCS
theory.\cite{Gubser81,Murata82,Green82,Brusetti82}  

Nevertheless, soon after the discovery of organic superconductivity, the
high-sensitivity of the superconducting state to
ir\-ra\-dia\-tion\cite{Choi82,Bouffard82} led Abri\-ko\-sov\cite{Abrikosov83}
to suggest the possibility of an unconventional -- triplet -- pairing,
although the non-magnetic nature of the induced defects is
questionable.\cite{Jerome04} The sensitivity to non-magnetic impurities, and
thus the existence of unconventional pairing, was later on clearly established
by the suppression of the superconducting transition upon alloying
(TMTSF)$_2$ClO$_4$ with a very small concentration of ReO$_4$
anions.\cite{Coulon82,Tomic83} A recent study\cite{Joo04} of the alloy \\
(TMTSF)$_2$(ClO$_4$)$_x$(ReO$_4$)$_{1-x}$ -- with different cooling rates and
different values of $x$ -- has confirmed this in remarkable way by showing
that the transition temperature $T_c$ is related to the scattering rate
$1/\tau$ by 
\beq
\ln \left(\frac{T_{c0}}{T_c}\right) = \Psi\left(\half + \frac{1}{4\pi\tau T_c}
\right) - \Psi\left(\half\right) 
\eeq
($T_{c0}$ is the transition temperature of the pure system and $\Psi$ the
digamma function), as expected for an unconventional superconductor in the
presence of non-magnetic impurities.\cite{Yuan03} 

Another indication of a possible unconventional pairing came from the
observation of Gor'kov and J\'erome\cite{Gorkov85} that the upper critical
field $H_{c2}(T)$, extrapolated down to $T=0$, would exceed the Pauli limited
field\cite{Clogston62,Chandrasekhar62} $H_P=1.84 T_{c0} /\mu_B\sim 2$ T by a
factor of 2. (The value of $H_P$ quoted here corresponds to $s$-wave pairing.) 
As spin-orbit interaction is weak in these systems and cannot provide an
explanation for such a large $H_{c2}$, it is tempting to again invoke
triplet pairing. This issue has been revived by recent measurements of the
upper critical field in (TMTSF)$_2$PF$_6$ with 
substantially improved accuracy in angular alignment and lower temperatures. 
Lee {\it et al.}\cite{Lee97,Lee00} observed a pronounced upward curvature
of $H_{c2}(T)$ without saturation -- down to $T\sim T_c/60$ -- for a field 
parallel to the $a$ or $b'$ axis, with $H^{b'}_{c2}(T)$ and $H^a_{c2}(T)$
exceeding the Pauli limited field $H_P$  by a factor of 4. Moreover,  
$H^{b'}_{c2}(T)$ becomes larger than $H^a_{c2}(T)$ at low temperatures. 
Similar results were
obtained from simultaneous resistivity and torque magnetization experiments in
(TMTSF)$_2$ClO$_4$.\cite{Oh04} The extrapolated value to zero temperature,
$H_{c2}(0)\sim 5$ T, is at least twice the Pauli limited field. 
 
\begin{figure}
\centerline{\includegraphics[bb=20 18 290 215,width=8cm]{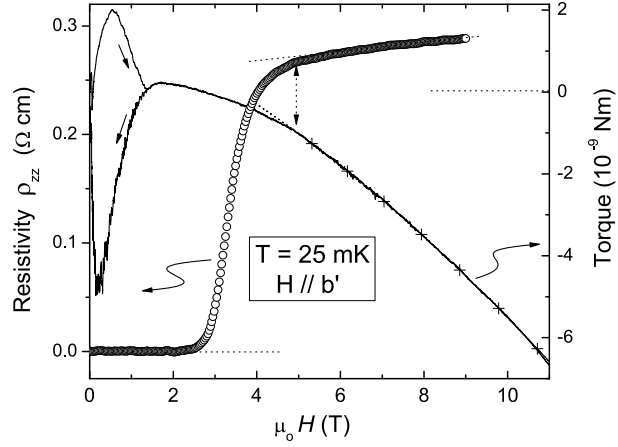}}
\caption{
 Resistivity (left scale) and torque magnetization (right) in
 (TMTSF)$_2$ClO$_4$ at 25 mK for $H\parallel b^\prime$. The dotted line and +
 symbols on the torque curve represent a temperature-independent normal state
 contribution. The onsets of diamagnetism and decreasing resistivity, upon
 decreasing field, are indicated by the arrow near $H_{c2}\sim$ 5T.  Arrows in
 the low field vortex state indicate field sweep directions. (Reprinted with
 permission from Ref.~\onlinecite{Oh04}. Copyright 2004 by the American
 Physical Society.) }
\end{figure}

There are different mechanisms that can greatly increase the orbital critical
field $H_{c2}^{\rm orb}(T)$ in organic conductors. Superconductivity in a
weakly-coupled plane system can survive in a strong parallel magnetic field if
the interplane (zero-field) coherence length $\xi_\perp(T)$ becomes smaller
than the interplane spacing $d$ at low temperature. Vortex cores, with size
$\xi_\perp(T)\lesssim d$, can then fit between planes without destroying the
superconducting order in the planes, and lead to a Josephson vortex
lattice. In the Bechgaard salts, even for a field parallel to the $b'$ axis,
the Josephson limit $\xi_\perp(T)\lesssim d$ is however unlikely to be
reached, since the interchain hopping amplitude $t_{\perp c}\sim 5-10$ K is
larger than the transition temperature $T_c\sim 1.1$ K. Nevertheless the
orbital critical field can be enhanced by a field-induced dimensional
crossover.\cite{Lebed86,Burlachkov87,Dupuis93,Dupuis94,Dupuis95} A magnetic
field parallel to the $b'$ axis tends to localize the wavefunctions in the
$(ac)$ planes, which in turn weakens the orbital destruction of the
superconducting order. When $\omega_c=eHc\gtrsim t_{\perp c}$ (which
corresponds to a field of a few Tesla in the Bechgaard salts), the wave
functions are essentially confined in the $(ac)$ planes and the orbital effect
of the field is completely suppressed. The coexistence between SDW and
superconductivity, as observed in a narrow pressure domain of the order of 0.8
kbar below the critical pressure $P_c$ (Fig.~\ref{fig:dia_gene}), can also
lead to a large increase of the orbital upper critical
field.\cite{Green80,Brusetti82a,Vuletic02,Lee02a,Lee05}   

Regardless of the origin of the large orbital critical field, another
mechanism is required to exceed the Pauli limited field $H_P$ in the Bechgaard
salts. For singlet spin pairing, the Pauli limit may be overcome by a
non-uniform 
Larkin-Ovchinnikov-Fulde-Ferrell (LOFF) state, where Cooper pairs form with a 
nonzero total momentum.\cite{Larkin65,Fulde64} This mechanism is particularly
efficient in a 1D system,\cite{Buzdin83,Lebed86,Dupuis93,Dupuis95}
due to the large phase space available for pairing at nonzero total momentum. 
For a linearized dispersion law, the mean-field upper critical field $H_c^{\rm
LOFF}$ diverges as $1/T$ in a pure superconductor. Lebed\cite{Lebed99} has
argued that the quasi-1D 
anisotropy reduces $H_c^{\rm LOFF}$ below the experimental observations. The
only possible explanation for a large upper critical field would then be an
equal-spin triplet pairing. A $p_x$-wave triplet state with a ${\bf d}$ vector
perpendicular to the $b'$ axis was proposed\cite{Lebed00} as a possible
explanation of the experimental observations reported in
Refs.~\onlinecite{Lee97,Lee00}.  

The triplet scenario in the Bechgaard salts is 
supported by recent NMR Knight shift experiments.\cite{Lee02,Lee03} Early NMR
experiments by Takigawa {\it et al.} already pointed to the unconventional
nature of the superconducting state in (TMTSF)$_2$ClO$_4$.\cite{Takigawa87}
The proton spin lattice relaxation rate $1/T_1$ does not exhibit a
Hebel-Slichter peak. It decreases rapidly just below $T_c$ in contrast to 
the typical BCS superconductor where it increases below $T_c$, reaching a
maximum at $T\sim 0.9 T_c$. Furthermore, $1/T_1\sim T^3$ for $T_c/2\lesssim
T \leq T_c$ -- as it is the case for most unconventional superconductors -- 
suggesting zeros or lines of zeros in the excitation spectrum. 
Recent experiments by Lee {\it et al.} in (TMTSF)$_2$PF$_6$ show that the
Knight shift, and therefore the spin susceptibility, remains unchanged at the
superconducting transition.\cite{Lee02,Lee03} This indicates triplet spin
pairing, since a singlet pairing would inevitably lead to a strong reduction
of the spin susceptibility ($\chi(T\to 0)\to 0$). It should however be noticed
that the interpretation of the Knight shift results -- due to a possible lack
of sample thermalization during the time of the experiment -- has been
questioned.\cite{Jerome04,Jerome05} 

\begin{figure}
\centerline{\includegraphics[width=8cm]{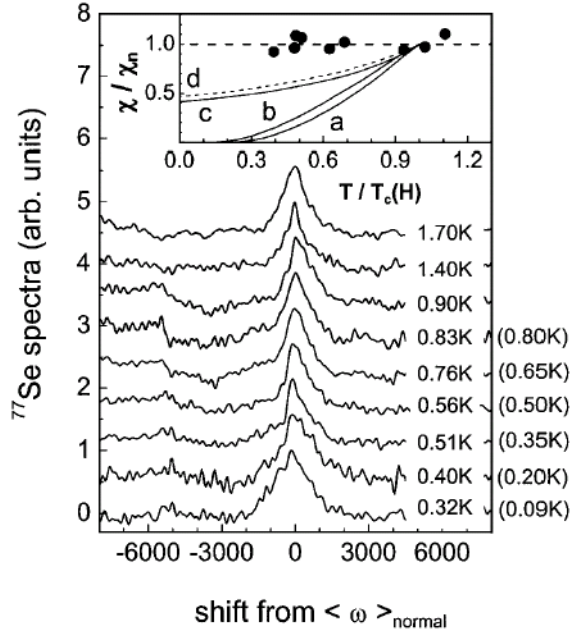}}
\caption{$^{77}$Se NMR spectra collected above and below $T_c$ (0.81 K at 1.43
  T). Each trace is normalized and offset for clarity. The temperatures shown
  in parentheses are the measured equilibrium temperatures before the
  pulse. In the inset, the spin susceptibility normalized by the normal state
  $\chi/\chi_n$ from measured first moments are compared with theoretical
  calculations\cite{Fulde65} for $H/H_{c2}(0)\sim 0$ (curve $a$) and 0.63
  (curve $b$). Curves 
  $c$ and $d$ are obtained from the ratio of applied field (1.43 T) to the
  measured upper critical field $H_{c2}(T)$ at which the superconducting
  criteria ``onset'' and ``50\% transition'' have been used, respectively, to
  determine $H_{c2}(T)$. (Reprinted with permission from
  Ref.~\onlinecite{Lee02}. Copyright 2002 by the American Physical Society.)}
\end{figure}

In principle, 
the symmetry of the order parameter can be determined from tunneling
spectroscopy. Sign changes of the pairing potential around the Fermi surface
lead to zero-energy bound states in the superconducting gap. These states
manifest themselves as a zero-bias peak in the tunneling conductance into the
corresponding edge.\cite{Sengupta01} More generally, different pairing
symmetries can be unambiguously distinguished by tunneling spectroscopy in a
magnetic field.\cite{Tanuma01,Tanuma02,Tanuma03} In practice however, the
realization of tunnel junctions with the TMTSF salts appears to be very
difficult. A large zero-bias conductance peak -- suggesting $p$-wave symmetry
-- across the junction between two organic superconductors was
observed.\cite{Ha03} 
But the absence of temperature broadening could indicate
that this peak is due to disorder rather than to a midgap 
state.\cite{Naughton05}

\begin{figure}
\centerline{\includegraphics[bb=90 510 430 770,width=8cm]{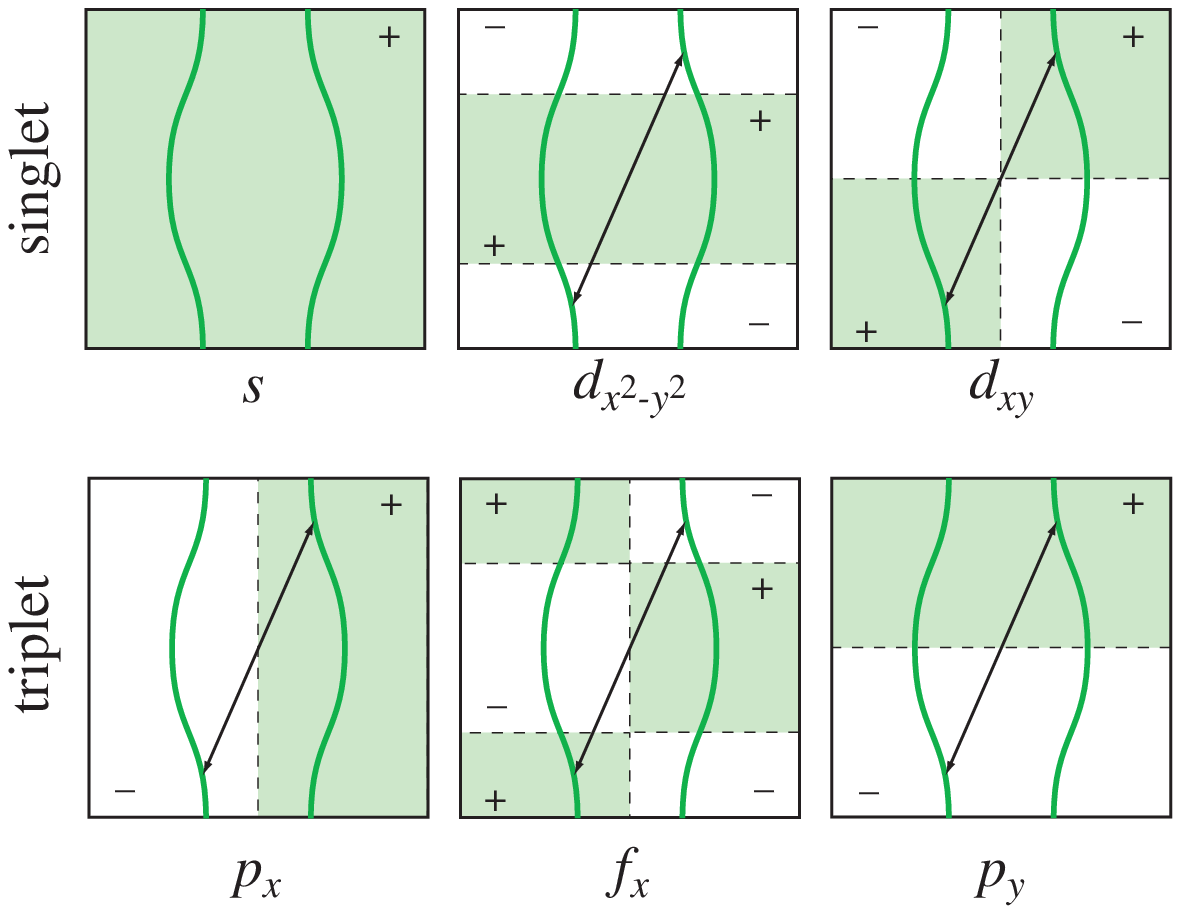}}
\caption{(color online) Gap symmetries $\Delta_r(\kperp)$ in a quasi-1D
  superconductor (after Ref.~\onlinecite{Fuseya05}, courtesy of Y. Suzumura). 
  $r=+/-$ denotes the right/left sheet of the Fermi surface.  
  (Singlet pairing) $s$: const, $d_{x^2-y^2}$: $\cos\kperp$,
  $d_{xy}$: $r\sin\kperp$. (Triplet pairing) $p_x$: $r$, $f$: $r\cos\kperp$,
  $p_y$: $\sin\kperp$. Next-nearest-neighbor and longer-range pairings are not
  considered. }
\label{fig:gap} 
\end{figure}

Information about the symmetry of the order parameter can also be obtained from
thermal conductivity measurements. The latter indicate the absence of nodes in
the excitation spectrum of the superconducting state in
(TMTSF)$_2$ClO$_4$,\cite{Belin97} 
thus suggesting a $p_x$-wave symmetry. However, because of the doubling of the
Fermi surface in the presence of anion ordering, a singlet $d$- or triplet
$f$-wave order phase would also be nodeless in (TMTSF)$_2$ClO$_4$ (see
Fig.~\ref{fig:gap} for the different gap symmetries in a quasi-1D 
superconductor).\cite{Bourbonnais99,Shimahara00} 

\section{MICROSCOPIC THEORIES OF THE SUPERCONDUCTING STATE}

\begin{figure}
\centerline{\includegraphics[width=6cm]{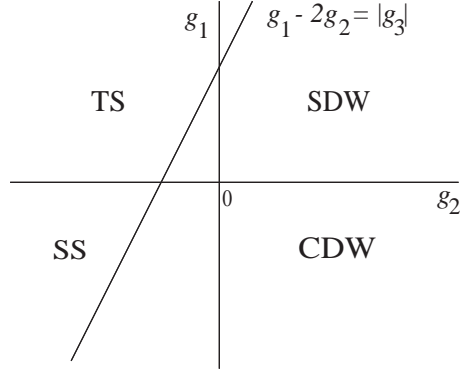}} 
\caption{Phase diagram (leading fluctuations) of the 1D electron gas
  in presence of Umklapp scattering. SS (TS):
  singlet (triplet) superconductivity.
  A gap develops in the charge sector (Mott insulating behavior) for
  $g_1-2g_2<|g_3|$.} 
\label{fig:dia_gology}
\end{figure}

The phase diagram of the 1D electron gas within the g-ology
frame\-work\cite{Solyom79} is shown in Fig.~\ref{fig:dia_gology}. $g_1$ and
$g_2$ denote the backward and forward scattering amplitudes, respectively, and
$g_3$ the strength of the (half-filling) Umklapp processes. Given the
importance of spin fluctuations in the phase diagram of the Bechgaard/Fabre
salts, as well as the existence of AF ground states, the
Bechgaard/Fabre salts should pertain to the upper right corner of the 1D phase
diagram ($g_1,g_2>0$ and $g_1-2g_2<|g_3|$) where the Umklapp processes are
relevant and the dominant fluctuations antiferromagnetic. In the Fabre salts, 
the non-magnetic insulating phase observed below $T_\rho \sim 100$ K indicates
the importance of Umklapp scattering and suggests sizable values  of
$g_3$ for this series. Since the long-range Coulomb interaction favors
$g_1<g_2$, the Fabre 
salts are expected to lie to the right of the phase diagram, i.e. far away from
the boundary $g_1-2g_2=|g_3|$. Since the triplet
superconducting phase is lying next to the SDW phase
(Fig.~\ref{fig:dia_gology}), it is tempting to invoke a 
change of the couplings $g_i$ under pressure to argue in favor of a $p_x$-wave
triplet superconducting state.\cite{Abrikosov83,Emery83} Such a drastic change
of the couplings, which would explain why (TMTTF)$_2$PF$_6$ becomes
superconducting above 4.35 GPa,\cite{Jaccard01,Wilhelm01,Adachi00} is however
somewhat unrealistic and has not received any theoretical backing so far. 
The Umklapp scattering being much weaker in the Bechgaard
salts, one cannot exclude that these compounds lie closer to the boundary
between the SDW and the triplet superconducting phase. A moderate change of the
couplings under pressure would then be sufficient to explain the
superconducting phase of (TMTSF)$_2$PF$_6$ observed above 6 kbar or
so. However, the destruction of the superconducting phase by a weak magnetic
field and the observation of a cascade of SDW phases for slightly higher
fields\cite{Chaikin96,Lederer96,Yakovenko96} would imply that the interaction
is strongly magnetic-field dependent -- again a very unlikely scenario. 

\begin{figure}
\centerline{\includegraphics[width=7cm]{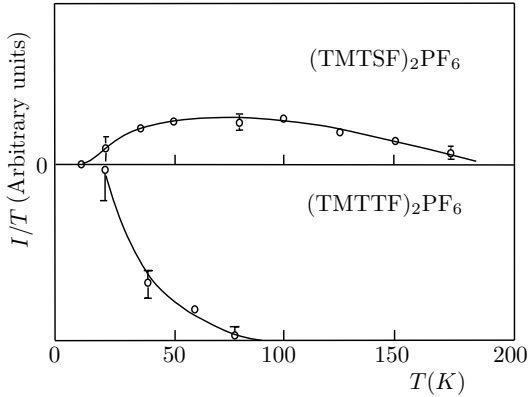}}
\caption{Temperature dependence of the $2k_F$ lattice susceptibility $(I/T)$
  as a function of temperature in the normal phase of (TMTSF)$_2$PF$_6$
  (top) and (TMTTF)$_2$PF$_6$ (bottom). (Reprinted with permission from
  Ref.~\onlinecite{Pouget96}. Copyright 1996 by EDP Sciences.) }
\label{fig:Xray}
\end{figure}

In all probability, the very origin of the superconducting
instability lies in the 3D behavior of these quasi-1D conductors. Thus the
attractive interaction is a consequence of a low-energy mechanism that
becomes more effective below the dimensional crossover temperature $T_x$.
Transverse hopping makes retarded electron-phonon interactions more effective,
since it is easier for the electrons to avoid the Coulomb
repulsion.\cite{Emery83} By comparing the sulfur and selenide series, it can
however be argued that, in the pressure range where superconductivity is
observed, the strength of the electron-phonon interaction is too weak to
explain the origin of the attractive interaction. For narrow tight-binding
bands in the organics, the attraction is strongest for backscattering
processes in which $2k_F$ phonons are
exchanged.\cite{Barisic70,Su79} According to the results of 
X-ray experiments performed on (TMTSF)$_2$X, however, the electron-phonon
vertex at this wave vector does not undergo any significant increase in
the normal state (Fig.~\ref{fig:Xray}). The amplitude of the $2k_F$ lattice
susceptibility in (TMTSF)$_2$PF$_6$ --
which is directly involved in the strength of the phonon exchange -- is
weak. It is instructive to compare with the sulfur analog compound
(TMTTF)$_2$PF$_6$, for which the electron-phonon vertex at $2k_F$ becomes
singular, signaling a lattice instability towards a spin-Peierls distortion
(Fig.~\ref{fig:Xray}). This instability produces a spin gap that is clearly
visible in the temperature dependence of the magnetic susceptibility and
nuclear relaxation rate.\cite{Creuzet87,Bourbonnais96} These effects are not
seen in (TMTSF)$_2$PF$_6$ close to $P_c$. The persistent enhancement of these
quantities indicates that interactions are dominantly repulsive
(Sec.~\ref{subsubsec:np}), making the traditional phonon-mediated source of
pairing inoperant.   

Emery\cite{Emery86} pointed out that near an SDW instability, short-range AF
spin fluctuations can give rise to anisotropic pairing and thus provide a
possible explanation of the origin of the superconducting phase in the
Bechgaard salts. Such fluctuations give rise to an oscillating
potential that couples to the electrons. Carriers can avoid the local Coulomb
repulsion and take advantage of the attractive part of this potential by moving
on different chains. This mechanism, which can lead to superconductivity at
low temperatures, is the spin-analog of the so-called Kohn-Luttinger mechanism
which assumes the pairing to originate in the exchange of charge-density
excitations produced by Friedel oscillations.\cite{Kohn65} While most
theoretical results on the spin-fluctuation-induced superconductivity are
based on RPA-like calculations,\cite{Beal86,Caron86,Bourbonnais88,Scalapino86,Scalapino87,Miyake86,Shimahara88,Kino99,Kuroki99,Scalapino95}  
the existence of such an electronic pairing mechanism in a quasi-1D conductor
has been recently confirmed by an RG approach.\cite{Duprat01} Moreover, it has
been recently realized that CDW fluctuations can play an important role in
stabilizing a triplet
phase.\cite{Kuroki01,Onari04,Tanaka04,Fuseya05,Bourbonnais04,Nickel05a,Nickel05b}
Below we discuss in simple terms the link between spin/charge 
fluctuations and unconventional pairing,\cite{Scalapino95} and present recent
results obtained from an RG approach.\cite{Bourbonnais04,Nickel05a,Nickel05b}

\subsection{Superconductivity from spin and charge fluctuations} 

Considering for the time being only intrachain
interactions, the interacting part of the Hamiltonian within the g-ology
framework\cite{Solyom79} reads
\beq
H_{\rm int} = \sum_{\q} [ 
g_{\rm ch} \rho(-\q)\rho(\q) + g_{\rm sp} \S(-\q)\cdot \S(\q) ] 
\label{Hint}
\eeq
(from now on we neglect the $c$ axis and consider a 2D model), 
where $\rho_\q$ and $\S_\q$ are the charge- and spin-density operators in the
Peierls channel ($q_x\sim 2k_F$), $g_{\rm ch}=g_1-g_2/2$ and $g_{\rm
  sp}=-g_2/2$. Starting from a half-filled extended Hubbard
model, we obtain $g_1=U-2V$ and $g_2=U+2V$, where $U$ is the onsite and $V$
the nearest-neighbor lattice site (dimer) interaction. For simplicity, we do
not consider Umklapp scattering ($g_3$), since it does not play an important
role in the present qualitative discussion. 
For repulsive interactions $g_1\sim g_2>0$, short-range spin fluctuations
develop at low temperatures due to the nesting of the Fermi surface. They can
be described by an  
effective Hamiltonian $H^{\rm eff}_{\rm int}$ obtained from (\ref{Hint}) by
replacing the bare coupling constants by their (static) RPA values
\bleq
g_{\rm ch}^{\rm RPA}(\q) &=& \frac{g_{\rm ch}}{1+g_{\rm ch}\chi_0(\q)} 
= g_{\rm ch} - g_{\rm ch}^2 \chi_{\rm ch}^{\rm RPA}(\q) ,
\nonumber \\ 
g_{\rm sp}^{\rm RPA}(\q) &=& \frac{g_{\rm sp}}{1+g_{\rm sp}\chi_0(\q)}  
= g_{\rm sp} - g_{\rm sp}^2 \chi_{\rm sp}^{\rm RPA}(\q) ,
\eleq
where $\chi^{\rm RPA}$ is the static ($\omega=0$) RPA susceptibility. The bare
particle-hole susceptibility diverges at low temperatures, i.e. $\chi_0(\Q)
\sim\ln(E_0/{\rm max}(T,t_{\perp b}'))$, due to the $\Q=(2k_F,\pi)$
nesting of the quasi-1D Fermi surface ($\epsilon_\k-\mu \simeq
-\epsilon_{\k+\Q}+\mu$).  ($E_0$ is a high-energy cutoff of the order of the
bandwidth.) The divergence is cut off by deviations from 
perfect nesting, characterized by the energy scale $t_{\perp b}'$
[Eq.~(\ref{dispersion})]. In the Bechgaard salts $t_{\perp b}'\sim 10$ K and
varies with pressure.  

\begin{figure}
\centerline{\includegraphics[width=6.5cm]{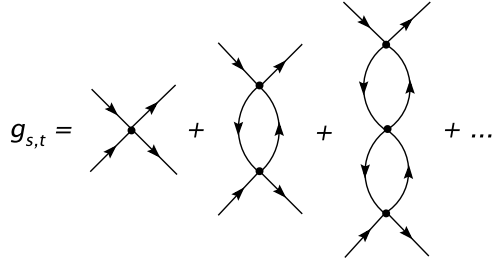}}
\caption{Diagrammatic representation of the effective interaction $g_{s,t}$ in
  the Cooper channel within the RPA. }
\label{fig:geff}
\end{figure}

When the nesting of the Fermi surface is good (small $t_{\perp b}'$), the spin
susceptibility $\chi_{\rm sp}^{\rm RPA}(\Q)$  diverges at low  temperatures,
thus signaling the formation of an SDW. A larger value of $t_{\perp b}'$
frustrates antiferromagnetism 
and, when exceeding a threshold value, eliminates the transition to the
SDW phase.\cite{Hasegawa86,Montambaux88} In that case, the (remaining)
short-range spin fluctuations can lead to pairing between fermions. To see
this, we rewrite the effective Hamiltonian $H^{\rm eff}_{\rm int}$ in the
particle-particle (Cooper) channel 
\beq
H_{\rm int}^{\rm eff} = \sum_{\k,\k'} [ g_s(\k,\k') O^*_s(\k) O_s(\k')  
+ g_t(\k,\k') \O^*_t(\k) \cdot \O_t(\k') ]
\label{Heff2}
\eeq
(we consider only Cooper pairs with zero total momentum), where 
\bleq
g_s(\k,\k') &=& - 3 g_{\rm sp}^{\rm RPA}(\k+\k') + g_{\rm ch}^{\rm
  RPA}(\k+\k') , \nonumber \\ 
g_t(\k,\k') &=&  - g_{\rm sp}^{\rm RPA}(\k+\k') - g_{\rm ch}^{\rm
  RPA}(\k+\k') 
\label{gst} 
\eleq
are the effective interactions in the singlet and triplet spin pairing
channels (Fig.~\ref{fig:geff}). $O_s(\k)$ ($\O_t(\k)$) is the annihilation
operator of a pair $(\k,-\k)$ in a singlet (triplet) spin state, and
$\O_t=(O_t^1,O_t^0,O_t^{-1})$ denotes the three components $S^z=1,0,-1$ of the
triplet state (total spin $S=1$). 

On the basis of the effective Hamiltonian (\ref{Heff2}) the BCS
theory predicts a superconducting transition whenever the effective
interaction $g_{s,t}$ turns out to be attractive in (at least) one pairing
channel. A simple argument shows that this is indeed always the case in the
presence of short-range spin fluctuations. The spin susceptibility $\chi^{\rm
  RPA}_{\rm sp}(\k+\k')$ exhibits a pronounced peak around
$\k+\k'=\Q$. Neglecting the unimportant $k_\para$ dependence, its Fourier
series expansion reads
\begin{multline}
\chi^{\rm RPA}_{\rm sp}(2k_F,\kperp+\kperpp) = \sum_{n=0}^\infty a_n (-1)^n
\cos[n(\kperp+\kperpp)]  \\
= \sum_{n=0}^\infty a_n (-1)^n [ \cos n\kperp \cos n\kperpp  -
  \sin n\kperp \sin n\kperpp ] , 
\label{chi}
\end{multline}
where $a_n\geq 0$. Choosing $a_n=a_0$, one obtains a diverging spin
susceptibility $\chi^{\rm RPA}_{\rm sp}(2k_F,\kperp+\kperpp) \propto 
\delta(\kperp+\kperpp-\pi)$. The condition $a_0>a_1>\cdots \geq 0$ gives a
broadened peak around $\kperp+\kperpp=\pi$. Eqs.~(\ref{gst},\ref{chi}) show
that the effective interaction in the singlet channel contains attractive
interactions for any value of $n$. In real space, $n$ corresponds to the range
of the pairing interaction in the $b$ direction. The dominant attractive
interaction corresponds to 
nearest-neighbor-chain pairing ($n=1$) and a $d_{x^2-y^2}$-wave order
parameter $\Delta_r(\kperp)\sim \cos\kperp$ ($r={\rm sgn}(k_x)$). The
interaction is also attractive in the triplet $f$-wave channel
($\Delta_r(\kperp)\sim r\cos\kperp$).  
However, all the three components of a (spin-one boson) SDW fluctuation
contribute to the superconducting coupling in the singlet channel -- hence the
factor of 3 in the first of equations (\ref{gst}). The latter therefore always
dominates over the triplet one when charge fluctuations are not important. 
Note that the interaction is repulsive in the singlet $d_{xy}$-wave
($\Delta_r(\kperp)\sim r\sin\kperp$) and the triplet $p_y$-wave ($\sin\kperp$)
channels. 

Eqs.~(\ref{gst}) show that CDW fluctuations tend to suppress the singlet
pairing, but reinforce the triplet one. In the Bechgaard salts, the physical
relevance of CDW fluctuations has been borne out by the puzzling observation
of a CDW that coexists with the SDW
(Sec.~\ref{subsec:itinerant}).\cite{Pouget96,Kagoshima99,Cao96} 
Within the framework of an extended
anisotropic Hubbard model, recent RPA calculations have shown that the triplet
$f$-wave pairing can overcome the singlet $d_{x^2-y^2}$-wave pairing when the
intrachain interactions are chosen such as to boost the CDW fluctuations with
respect to the SDW ones.\cite{Kuroki01,Onari04,Tanaka04} In a half-filled
model, this however requires the nearest-neighbor (intrachain) interaction $V$
to exceed $U/2$. In a quarter-filled
model -- appropriate if one ignores the weak dimerization along the chains --
the condition for $f$-wave superconductivity becomes $V_2\geq U/2$ ($V_2$ is
the next-nearest-neighbor (intrachain) Coulomb interaction) and 
appears even more unrealistic. Similar conclusions were reached within an RG
approach.\cite{Fuseya05}

Given that electrons interact through the Coulomb interaction, not only
intrachain but also {\it inter}chain interactions are present in practice. At
large momentum transfer, the interchain interaction is well known to favor a
CDW ordered state.\cite{Gorkov74,Mihaly76,Lee77,Menyhard77} This mechanism is
mostly responsible for CDW long-range order observed in several organic and
inorganic low-dimensional solids (e.g. TTF-TCNQ).\cite{Barisic85,Pouget89} In
the Bechgaard salts, both the interchain Coulomb interaction and the kinetic
interchain coupling ($t_{\perp b}$) are likely to 
be important in the temperature range where superconductivity and SDW
instability occur, and should be considered on equal footing. An RG approach
has recently been used to determine the phase diagram of an extended quasi-1D
electron gas model that includes interchain hopping, nesting deviations and
both intrachain and interchain
interactions.\cite{Bourbonnais04,Nickel05a,Nickel05b}  The intrachain
interactions turn out to have a sizeable impact on the structure of the phase
diagram. Unexpectedly, for reasonably small values of the interchain
interactions, the singlet $d_{x^2-y^2}$-wave superconducting phase is
destabilized to the 
benefit of the triplet $f$-wave phase with a similar range of $T_c$. The SDW
phase is also found to be close in stability to a CDW phase. Before
presenting these results in more detail (Sec.~\ref{subsubsec:rg}), let us
discuss in simple terms the role of interchain interactions. The interchain
backward scattering amplitude $g_1^\perp$ ($>0$) contributes to the
effective interaction in the Cooper channel,
\bleq
g_s(\kperp,\kperpp) &\to& g_s(\kperp,\kperpp) + 2 g_1^\perp
[\cos\kperp\cos\kperpp - \sin\kperp\sin\kperpp ] , \nonumber \\ 
g_t(\kperp,\kperpp) &\to& g_t(\kperp,\kperpp) + 2 g_1^\perp
[-\cos\kperp\cos\kperpp + \sin\kperp\sin\kperpp ] .
\eleq 
It thus tends to suppress singlet $d_{x^2-y^2}$ pairing, but favors triplet
$f$-wave pairing. In addition to this ``direct'' contribution, $g_1^\perp$
reinforces CDW fluctuations,
\beq 
g_{\rm ch}(\qperp) \to g_{\rm ch}(\qperp) + 2g_1^\perp \cos\qperp ,
\eeq 
and therefore enhances the $f$-wave pairing over the $d_{x^2-y^2}$-wave
pairing  {\it via} the mechanism of fluctuation exchange [see
  Eq.~(\ref{gst})]. As for the interchain forward scattering $g_2^\perp$, its
direct contribution to the DW channel is negligible, but it has a detrimental
effect on both singlet and triplet nearest-neighbor-chain 
pairings. This latter effect, which is neutralized by the Umklapp
scattering processes, can lead to next-nearest-neighbor-chain pairings 
when Umklapp processes are very weak.\cite{Nickel05b}  

\subsection{RG calculation of the phase diagram of quasi-1D conductors} 
\label{subsubsec:rg}

\begin{figure}
\centerline{\includegraphics[bb=135 545 353 678,width=7.5cm]{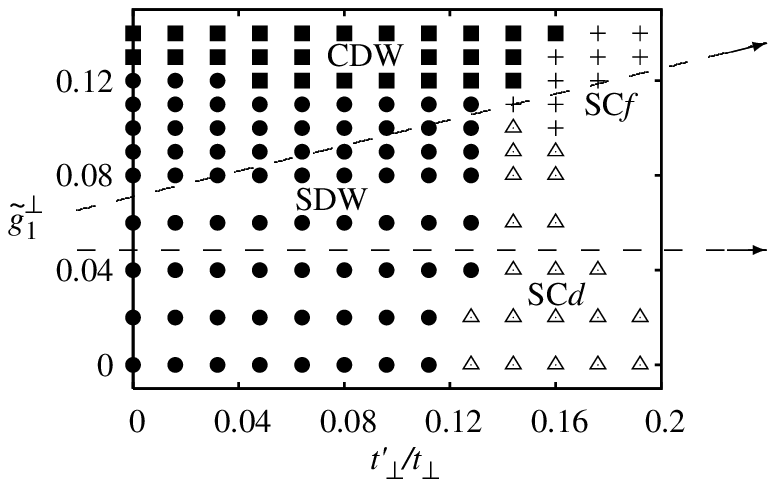}}
\caption{$T=0$ phase diagram as a function of $t'_\perp/t_\perp$ and
  $\tilde g^\perp_1$. Circles: SDW, squares: CDW, triangles: SC$d$ 
($\Delta_r(k_\perp)\propto\cos k_\perp$), crosses: SC$f$
  ($\Delta_r(k_\perp) \propto r \cos k_\perp$). The dashed lines indicate
  two (among many) possible pressure axes, corresponding to transitions
  SDW$\to$SC$d$ and SDW$\to$SC$f$.\cite{Nickel05a,Nickel05b} }
\label{fig:diarg1}  
\end{figure}

As a systematic and unbiased method with no {\it a priori} assumption, 
the RG  method is perfectly suited to study competing instabilities. The
zero-temperature phase diagram obtained with this technique is shown in
Fig.~\ref{fig:diarg1}.\cite{Nickel05a,Nickel05b} In the absence
of interchain 
interactions ($g_1^\perp=g_2^\perp=0$), it confirms the validity of the
qualitative arguments given above. When the nesting of the Fermi surface is
nearly perfect (small $t_\perp'$) the ground state is an SDW. Above a
threshold value of $t_\perp'$, the low-temperature SDW instability is
suppressed and the ground state becomes a $d_{x^2-y^2}$-wave superconducting
(SC$d$) state with an order parameter $\Delta_r(k_\perp)\propto \cos
k_\perp$.\cite{Duprat01} In the presence of interchain interactions
($g_1^\perp>0$), the region of stability of the SC$d$ phase shrinks, and a
triplet superconducting $f$-wave (SC$f$) phase appears next to the $d$-wave
phase for $\tilde g_1^\perp=g_1^\perp/\pi v_F \simeq 0.1$ -- obtained here for
typical values of intrachain couplings and band
parameters.\cite{Nickel05a,Nickel05b} 
For larger values of the interchain interactions, the
SC$d$ phase disappears and the region of stability of the $f$-wave
superconducting phase widens. In addition a CDW phase appears, thus giving the
sequence of phase transitions SDW$\to$CDW$\to$SC$f$ as a function of
$t'_\perp$. For $\tilde g^\perp_1 \gtrsim 0.12$, the SDW phase disappears.  
Note that for $\tilde g^\perp_1\simeq 0.11$, the region of
stability of the CDW phase is very narrow, and there is essentially a direct
transition between the SDW and SC$f$ phases.

\begin{figure}  
\centerline{\includegraphics[bb=50 65 410 300,width=7cm]{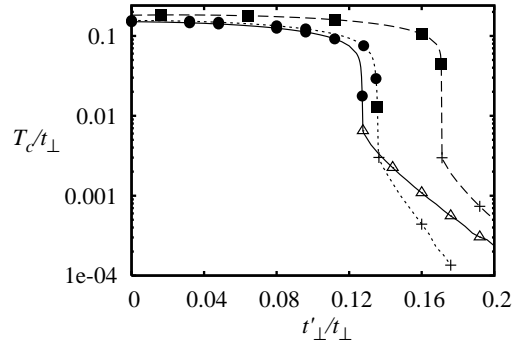}}
\caption{Transition temperature as a function of $t'_\perp/t_\perp$ for
  $\tilde g^\perp_1=0$, 0.11 and 0.14, corresponding to solid, dotted, and
  dashed lines, respectively.\cite{Nickel05a,Nickel05b} }  
\label{fig:diarg2}
\end{figure}

The RG calculations yield $T_c\sim 30$ K for the SDW phase in the
case of perfect nesting and 
$T_c\sim 0.6-1.2$ K for the superconducting phase, in reasonable
agreement with the experimental observations in the Bechgaard salts. 
Fig.~\ref{fig:diarg2} shows the  
transition temperature $T_c$ as a function of $t'_\perp$ for three different
values of the interchain interactions, $\tilde g^\perp_1=0$, 0.11 and 0.14,
corresponding to the three different sequences of phase transitions as a
function of $t_\perp'$: SDW$\to$SC$d$, SDW$\to$(CDW)$\to$SC$f$ and
CDW$\to$SC$f$. The phase diagram is unchanged when both $g_2^\perp$ and a weak
Umklapp scattering amplitude $g_3$ are included.\cite{Nickel05a,Nickel05b} 

\begin{figure}  
\centerline{\includegraphics[bb=50 65 410 300,width=6.5cm]{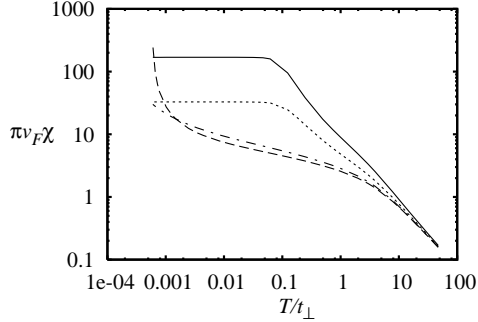}}  
\caption{Temperature dependence of the susceptibilities in the normal phase
  above the SC$d$ phase ($t'_\perp=0.152 t_\perp$ and
  $\tilde g_1^\perp=0.08$). The continuous, dotted, dashed, and dashed-dotted
  lines correspond to SDW, CDW, SC$d$ and SC$f$ correlations,
  respectively.\cite{Nickel05a,Nickel05b} }  
\label{fig3}
\end{figure}

\begin{figure}   
\centerline{\includegraphics[bb=94 588 420 695,width=12.4cm]{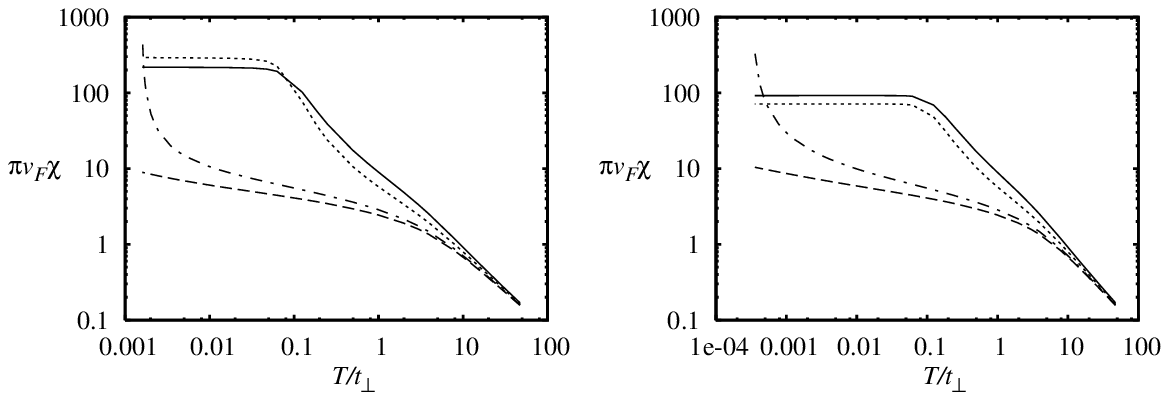}}
\caption{Temperature dependence of the susceptibilities in the normal phase
  above the SC$f$ phase for $\tilde g^\perp_1=0.12$, $t_\perp'=0.152 t_\perp$
  (left) and $t'_\perp=0.176t_\perp$ (right).\cite{Nickel05a,Nickel05b} }  
\label{fig4} 
\end{figure}

The RG approach also provides important information about the fluctuations in
the normal phase. The dominant fluctuations above the SC$d$ phase are SDW
fluctuations as observed experimentally (Sec.~\ref{subsec:itinerant}). 
Although they saturate below $T\sim t'_\perp$ where the 
SC$d$ fluctuations become more and more important, the latter
dominate only in a very narrow temperature range above the superconducting
transition (Fig.~\ref{fig3}). 
Above the SC$f$ and CDW phases, one expects strong CDW
fluctuations driven by $g^\perp_1$. Fig.~\ref{fig4} shows that for
$\tilde g^\perp_1 \sim 0.11-0.12$, strong SDW and CDW fluctuations coexist
above the SC$f$ 
phase. Remarkably, there are regions of the phase diagram where the
SDW fluctuations remain the dominant ones in the normal phase above the SC$f$
or CDW phase (right panel in Fig.~\ref{fig4}).

A central result of the RG calculation is the close proximity of SDW, CDW and
SC$f$ phases in the phase diagram of a quasi-1D conductor within a {\it
realistic} range of values for  the repulsive intrachain and interchain
interactions. Although this proximity is found only in a small range of
interchain interactions, there are several features that
suggest that this part of the phase diagram is the relevant one for the
Bechgaard salts. i)
SDW fluctuations remain important in the normal phase throughout the whole
phase diagram. They are the dominant fluctuations above the SC$d$ phase, and
remain strong -- being sometimes even dominant -- above the SC$f$ phase where
they coexist with strong CDW fluctuations, in accordance with
observations.\cite{Wzietek93,Cao96} ii) The SC$f$ and CDW phases stand 
nearby in the theoretical phase diagram, the CDW phase always closely
following the SC$f$ phase when the interchain interactions increase. This
agrees with the experimental finding that both SDW and CDW coexist in the DW
phase of the Bechgaard salts\cite{Pouget96,Kagoshima99} and the existence,
besides SDW correlations, of CDW fluctuations in the normal state above the  
superconducting phase.\cite{Cao96} iii) Depending how one moves in practice 
in the phase diagram as a function of pressure, these results are 
compatible with either a singlet $d_{x^2-y^2}$-wave or a triplet $f$-wave
superconducting phase in the Bechgaard salts (see the two pressure axes in
Fig.~\ref{fig:diarg1}). Moreover, one
cannot exclude that both SC$d$ and SC$f$ phases exist in these materials,
with the sequence of phase transitions SDW$\to$SC$d\to$SC$f$ as a
function of pressure. It is also possible that the SC$f$ phase is stabilized
by a magnetic field,\cite{Shimahara00b} since an equal-spin pairing
triplet phase is 
not sensitive to the Pauli pair breaking effect contrary to the SC$d$
phase. This would make possible the existence of large upper critical fields
exceeding the Pauli limit,\cite{Lee97,Oh04} and would also provide an
explanation for the temperature independence of the NMR Knight shift in the
superconducting phase.\cite{Lee02}

\section{CONCLUSION} 

Notwithstanding the recent experimental progresses, many of the basic
questions related to superconductivity in the Bechgaard and Fabre salts remain
largely open. The very nature of the superconducting state -- the orbital
symmetry of the order parameter and the singlet/triplet character of the
pairing -- is still not known without ambiguity even though recent upper
critical field measurements\cite{Lee97,Lee00,Oh04} and NMR
experiments\cite{Lee02,Lee03} support a triplet pairing.  

We argued that the conventional electron-phonon mechanism is unable to explain
the origin of the superconducting phase. On the other hand, the proximity of
the SDW phase, as well as the observation of strong spin fluctuations in the
normal state precursor to the superconducting
phase,\cite{Jaccard01,Wilhelm01,Wzietek93} strongly suggest an
intimate relationship between antiferromagnetism and superconductivity in the
Bechgaard/Fabre salts. The scenario originally proposed by Emery,\cite{Emery86}
whereby short-range AF spin fluctuations can give rise to anisotropic pairing
and thus stabilize a superconducting phase, is so far the only one that is
consistent with the experimental observations and the repulsive nature of the 
electron-electron interactions. 

Within the framework of the RG approach, it has recently been shown that 
when spin and charge fluctuations are taken into account on equal
footing, both singlet $d_{x^2-y^2}$- and triplet $f$-wave superconducting
phases can emerge at low temperatures whenever the nesting properties of the
Fermi surface deteriorate under
pressure.\cite{Kuroki01,Onari04,Tanaka04,Fuseya05,Bourbonnais04,Nickel05a,Nickel05b}  CDW fluctuations are enhanced by the
long-range part of the Coulomb interaction. Remarkably, for a reasonably small
value of the interchain interactions, the singlet $d_{x^2-y^2}$-wave
phase is destabilized to the benefit of a triplet $f$-wave with a similar range
of $T_c$.\cite{Nickel05a,Nickel05b} The physical relevance of CDW fluctuations
in the Bechgaard salts has been born out by the observation of a CDW that
actually coexists with the SDW.\cite{Pouget96,Kagoshima99} CDW fluctuations
were also observed in the normal state precursor to the superconducting
state.\cite{Cao96}  

As a systematic and unbiased method with no {\it a priori} assumptions, the RG
has proven to be a method of choice to study the physical properties of
quasi-1D organic conductors. An important theoretical issue is now to go
beyond the instabilities of the normal state. On the one hand, the RG analysis
should be extended to the low-temperature broken-symmetry
phases in order to study the possible coexistence of
superconductivity and antiferromagnetism, as well as CDW and SDW, as observed
in the Bechgaard salts.\cite{Vuletic02,Lee05,Pouget96,Kagoshima99} On the
other hand, the RG technique might also enable to tackle the unusual
properties of the metallic phase. A recent RG analysis\cite{Fuseya05}
of the AF spin susceptibility in the normal phase has shown that below the
dimensional crossover temperature, it differs significantly from the
prediction of single-channel (RPA) theories. The interplay between
the superconducting and Peierls channels, which is at the origin of
spin-fluctuation induced superconductivity, might also be responsible for the
unusual properties of the metallic state below the dimensional crossover
temperature. 


\begin{thebibliography}{100}

\bibitem{Jerome80}
D. J\'erome, A. Mazaud, M. Ribault, and K. Bechgaard, J. de Phys. (Paris) Lett.
  {\bf 41},  L95  (1980).

\bibitem{Jerome82}
D. J{\'e}rome and H.~J. Schulz, Adv. Phys. {\bf 31},  299  (1982).

\bibitem{Ishiguro90}
T. Ishiguro and K. Yamaji, {\em Organic Superconductors}, Vol.~88 of {\em
  Springer-Verlag Series in Solid-State Science} (Springer-Verlag, Berlin,
  Heidelberg, 1990).

\bibitem{McKenzie98}
R.~H. McKenzie, Comments Cond. Matt. Phys. {\bf 18},  309  (1998).

\bibitem{Lefebvre00}
S. Lefebvre {\it et~al.}, Phys. Rev. Lett. {\bf 85},  5420  (2000).

\bibitem{Flouquet05}
J. Flouquet, arXiv:cond-mat/0501602 (unpublished).

\bibitem{Jerome04}
D. J\'erome, Chem. Rev. {\bf 104},  5565  (2004).

\bibitem{Bechgaard80}
K. Bechgaard {\it et~al.}, Solid State Comm. {\bf 33},  1119  (1980).

\bibitem{Bourbonnais99}
C. Bourbonnais and D. J{\'e}rome,  in {\em Advances in Synthetic Metals, Twenty
  Years of Progress in Science and Technology}, edited by P. Bernier, S.
  Lefrant, and G. Bidan (Elsevier, New York, 1999), pp.\ 206--261,
  arXiv:cond-mat/9903101.

\bibitem{Grant83}
P.~M. Grant, J. de Phys. (Paris) Coll. {\bf 44},  847  (1983).

\bibitem{Yamaji82}
K. Yamaji, J. Phys. Soc. Jpn. {\bf 51},  2787  (1982).

\bibitem{Ducasse86}
L. Ducasse {\it et~al.}, J. Phys. C {\bf 19},  3805  (1986).

\bibitem{Balicas94}
L. Balicas {\it et~al.}, J. Phys. I {\bf 4},  1539  (1994).

\bibitem{Emery82}
V.~J. Emery, R. Bruisma, and S. Bari\v{s}i\'c, Phys. Rev. Lett. {\bf 48},  1039
   (1982).

\bibitem{Barisic81}
S. Bari\v{s}i\'c and S. Brazovskii,  in {\em Recent Developments in Condensed
  Matter Physics}, edited by J.~T. Devreese (Plenum, New York, 1981), Vol.~1,
  p.\ 327.

\bibitem{Jaccard01}
D. Jaccard {\it et~al.}, J. Phys.: Cond. Matt. {\bf 13},  L89  (2001).

\bibitem{Wilhelm01}
H. Wilhelm {\it et~al.}, Eur. Phys. J. B {\bf 21},  175  (2001).

\bibitem{Adachi00}
T. Adachi {\it et~al.}, J. Am. Chem. Soc. {\bf 122},  3238  (2000).

\bibitem{Bourbonnais98}
C. Bourbonnais and D. J{\'e}rome, Science {\bf 281},  1156  (1998).

\bibitem{Haldane81}
F.~D.~M. Haldane, J. Phys. C {\bf 14},  2585  (1981).

\bibitem{Giamarchi_book}
T. Giamarchi, {\em Quantum Physics in One Dimension} (Oxford University Press,
  Oxford, 2004).

\bibitem{Coulon82a}
C. Coulon {\it et~al.}, J. Phys. (Paris) {\bf 43},  1059  (1982).

\bibitem{Schwartz98}
A. Schwartz {\it et~al.}, Phys. Rev. B {\bf 58},  1261  (1998).

\bibitem{Wzietek93}
P. Wzietek {\it et~al.}, J. Phys. I {\bf 3},  171  (1993).

\bibitem{Bourbonnais89}
C. Bourbonnais {\it et~al.}, Phys. Rev. Lett. {\bf 62},  1532  (1989).

\bibitem{Bourbonnais93}
C. Bourbonnais, J. Phys. I {\bf 3},  143  (1993).

\bibitem{Bourbonnais88}
C. Bourbonnais and L.~G. Caron, Europhys. Lett. {\bf 5},  209  (1988).

\bibitem{Bourbonnais91}
C. Bourbonnais and L.~G. Caron, Int. J. Mod. Phys. B {\bf 5},  1033  (1991).

\bibitem{Brown97}
S.~E. Brown {\it et~al.}, Synth. Metals {\bf 86},  1937  (1997).

\bibitem{Bourbonnais84}
C. Bourbonnais, F. Creuzet, D. J\'erome, and K. Bechgaard, J. de Phys. (Paris)
  Lett. {\bf 45},  L755  (1984).

\bibitem{Jacobsen81}
C.~S. Jacobsen, D.~B. Tanner, and K. Bechgaard, Phys. Rev. Lett. {\bf 46},
  1142  (1981).

\bibitem{Jacobsen83}
C.~S. Jacobsen, D.~B. Tanner, and K. Bechgaard, Phys. Rev. B {\bf 28},  7019
  (1983).

\bibitem{Cao96}
N. Cao, T. Timusk, and K. Bechgaard, J. Phys. I {\bf 6},  1719  (1996).

\bibitem{Dressel96}
M. Dressel, A. Schwartz, G. Gr\"uner, and L. Degiorgi, Phys. Rev. Lett. {\bf
  77},  398  (1996).

\bibitem{Giamarchi97}
T. Giamarchi, Physica B {\bf 230-232},  975  (1997).

\bibitem{Vescoli98}
V. Vescoli {\it et~al.}, Science {\bf 281},  1181  (1998).

\bibitem{Giamarchi04}
T. Giamarchi, Chem. Rev. {\bf 104},  5037  (2004).

\bibitem{Favand96}
J. Favand and F. Mila, Phys. Rev. B {\bf 54},  10 425  (1996).

\bibitem{Moser00}
J. Moser {\it et~al.}, Phys. Rev. Lett. {\bf 84},  2674  (2000).

\bibitem{Auban99}
P. Auban-Senzier, D. J\'erome, and J. Moser,  in {\em Physical Phenomena at
  High Magnetic Fields}, edited by Z. Fisk, L. Gor'kov, and L. Schrieffer
  (World Scientific, Singapore, 1999).

\bibitem{Mihaly00}
G. Mih\'aly, I. K\'ezsm\'arki, F. Z\'amborszky, and L. Forro, Phys. Rev. Lett.
  {\bf 84},  2670  (2000).

\bibitem{Fertey99}
P. Fertey, M. Poirier, and P. Batail, Eur. Phys. J. B {\bf 10},  305  (1999).

\bibitem{Moser98}
J. Moser {\it et~al.}, Eur. Phys. J. B {\bf 1},  39  (1998).

\bibitem{Zwick97}
F. Zwick {\it et~al.}, Phys. Rev. Lett. {\bf 79},  3982  (1997).

\bibitem{Dardel93}
B. Dardel {\it et~al.}, Europhys. Lett. {\bf 24},  687  (1993).

\bibitem{Georges00}
A. Georges, T. Giamarchi, and N. Sandler, Phys. Rev. B {\bf 61},  16393
  (2000).

\bibitem{Danner94}
G.~M. Danner, W. Kang, and P.~M. Chaikin, Phys. Rev. Lett. {\bf 72},  3714
  (1994).

\bibitem{Gorkov96}
L.~P. Gor'kov, J. Phys. I {\bf 6},  1697  (1996).

\bibitem{Zheleznyak99}
A.~T. Zheleznyak and V.~M. Yakovenko, Eur. Phys. J. B {\bf 11},  385  (1999).

\bibitem{Chaikin96}
P.~M. Chaikin, J. Phys. I {\bf 6},  1875  (1996).

\bibitem{Lederer96}
P. Lederer, J. Phys. I {\bf 6},  1899  (1996).

\bibitem{Yakovenko96}
V.~M. Yakovenko and H.~S. Goan, J. Phys. I {\bf 6},  1917  (1996).

\bibitem{Pouget96}
J.~P. Pouget and S. Ravy, J. Phys. I {\bf 6},  1501  (1996).

\bibitem{Kagoshima99}
S. Kagoshima {\it et~al.}, Sol. State Comm. {\bf 110},  479  (1999).

\bibitem{Seo97}
H. Seo and H. Fukuyama, J. Phys. Soc. Jpn. {\bf 66},  1249  (1997).

\bibitem{Kobayashi98}
N. Kobayashi, M. Ogata, and K. Yonemitsu, J. Phys. Soc. Jpn. {\bf 67},  1098
  (1998).

\bibitem{Mazumdar99}
S. Mazumdar, S. Ramasesha, R.~T. Clay, and D.~K. Campbell, Phys. Rev. Lett.
  {\bf 82},  1522  (1999).

\bibitem{Tomio00}
Y. Tomio and Y. Suzumura, J. Phys. Soc. Jpn. {\bf 69},  796  (2000).

\bibitem{Tomio01}
Y. Tomio and Y. Suzumura, J. Phys. Chem. Solids {\bf 62},  431  (2001).

\bibitem{Garoche82}
P. Garoche, R. Brusetti, and K. Bechgaard, Phys. Rev. Lett. {\bf 49},  1346
  (1982).

\bibitem{Garoche82a}
P. Garoche, R. Brusetti, and D. J\'erome, J. de Phys. (Paris) Lett. {\bf 43},
  L147  (1982).

\bibitem{Gubser81}
D.~L. Gubser {\it et~al.}, Phys. Rev. B {\bf 24},  478  (1981).

\bibitem{Murata82}
K. Murata {\it et~al.}, Mol. Cryst. Liq. Cryst. {\bf 79},  639  (1982).

\bibitem{Green82}
R.~L. Green, P. Haen, S.~Z. Huang, and E.~M. Engler, Mol. Cryst. Liq. Cryst.
  {\bf 79},  183  (1982).

\bibitem{Brusetti82}
R. Brusetti, M. Ribault, D. J\'erome, and K. Bechgaard, J. Phys. (France) {\bf
  43},  52  (1982).

\bibitem{Choi82}
M.~Y. Choi {\it et~al.}, Phys. Rev. B {\bf 25},  6208  (1985).

\bibitem{Bouffard82}
S. Bouffard {\it et~al.}, J. Phys. C {\bf 15},  2951  (1982).

\bibitem{Abrikosov83}
A.~A. Abrikosov, JETP Lett. {\bf 37},  503  (1983).

\bibitem{Coulon82}
C. Coulon {\it et~al.}, J. Phys. (France) {\bf 43},  1721  (1982).

\bibitem{Tomic83}
S. Tomic {\it et~al.}, J. de Phys. (Paris) Coll. {\bf 44},  C3  (1983).

\bibitem{Joo04}
N. Joo {\it et~al.}, Eur. Phys. J. B {\bf 40},  43  (2004).

\bibitem{Yuan03}
Q. Yuan {\it et~al.}, Phys. Rev. B {\bf 68},  174510  (2003).

\bibitem{Gorkov85}
L.~P. Gor'kov and D. J\'erome, J. de Phys. (Paris) Lett. {\bf 46},  L643
  (1985).

\bibitem{Clogston62}
A.~M. Clogston, Phys. Rev. Lett. {\bf 9},  266  (1962).

\bibitem{Chandrasekhar62}
B.~S. Chandrasekhar, Appl. Phys. Lett. {\bf 1},  7  (1962).

\bibitem{Lee97}
I.~J. Lee, M.~J. Naughton, G.~M. Danner, and P.~M. Chaikin, Phys. Rev. Lett.
  {\bf 78},  3555  (1997).

\bibitem{Lee00}
I.~J. Lee, P.~M. Chaikin, and M.~J. Naughton, Phys. Rev. B {\bf 62},  R14669
  (2000).

\bibitem{Oh04}
J.~I. Oh and M.~J. Naughton, Phys. Rev. Lett. {\bf 92},  067001  (2004).

\bibitem{Lebed86}
A.~G. Lebed, JETP Lett. {\bf 44},  114  (1986).

\bibitem{Burlachkov87}
L.~I. Burlachkov, L.~P. Gor'kov, and A.~G. Lebed, Europhys. Lett. {\bf 4},  941
   (1987).

\bibitem{Dupuis93}
N. Dupuis, G. Montambaux, and C.~A. R.~S. de~Melo, Phys. Rev. Lett. {\bf 70},
  2613  (1993).

\bibitem{Dupuis94}
N. Dupuis and G. Montambaux, Phys. Rev. B {\bf 49},  8993  (1994).

\bibitem{Dupuis95}
N. Dupuis, Phys. Rev. B {\bf 51},  9074  (1995).

\bibitem{Green80}
R.~L. Greene and E.~M. Engler, Phys. Rev. Lett. {\bf 45},  1587  (1980).

\bibitem{Brusetti82a}
R. Brusetti, M. Ribault, D. J\'erome, and K. Bechgaard, J. Phys. (France) {\bf
  43},  801  (1982).

\bibitem{Vuletic02}
T. Vuleti\'c {\it et~al.}, Eur. Phys. J. B {\bf 25},  319  (2002).

\bibitem{Lee02a}
I.~J. Lee, P.~M. Chaikin, and M.~J. Naughton, Phys. Rev. Lett. {\bf 88},
  207002  (2002).

\bibitem{Lee05}
I.~J. Lee {\it et~al.}, Phys. Rev. Lett. {\bf 94},  197001  (2005).

\bibitem{Larkin65}
A.~I. Larkin and Y.~N. Ovchinnikov, Sov. Phys. JETP {\bf 20},  762  (1965).

\bibitem{Fulde64}
P. Fulde and R.~A. Ferrell, Phys. Rev. {\bf 135},  A550  (1965).

\bibitem{Buzdin83}
A.~I. Buzdin and S.~V. Polonskii, Sov. Phys. JETP {\bf 66},  422  (1983).

\bibitem{Lebed99}
A.~G. Lebed, Phys. Rev. B {\bf 59},  R721  (1999).

\bibitem{Lebed00}
A.~G. Lebed, K. Machida, and M. Ozaki, Phys. Rev. B {\bf 62},  R795  (2000).

\bibitem{Lee02}
I.~J. Lee {\it et~al.}, Phys. Rev. Lett. {\bf 88},  017004  (2002).

\bibitem{Lee03}
I.~J. Lee {\it et~al.}, Phys. Rev. B {\bf 68},  092510  (2003).

\bibitem{Takigawa87}
M. Takigawa, H. Yasuoka, and G. Saito, J. Phys. Soc. Jpn. {\bf 56},  873
  (1987).

\bibitem{Jerome05}
D. J\'erome and C.~R. Pasquier,  in {\em Superconductors}, edited by A.~V.
  Narlikar (Springer-Verlag, Berlin, 2005).

\bibitem{Fulde65}
P. Fulde and K. Maki, Phys. Rev. B {\bf 139},  A788  (1965).

\bibitem{Sengupta01}
K. Sengupta {\it et~al.}, Phys. Rev. B {\bf 63},  144531  (2001).

\bibitem{Tanuma01}
Y. Tanuma, K. Kuroki, Y. Tanaka, and S. Kashiwaya, Phys. Rev. B {\bf 64},
  214510  (2001).

\bibitem{Tanuma02}
Y. Tanuma {\it et~al.}, Phys. Rev. B {\bf 66},  094507  (2002).

\bibitem{Tanuma03}
Y. Tanuma, K. Kuroki, Y. Tanaka, and S. Kashiwaya, Phys. Rev. B {\bf 68},
  214513  (2003).

\bibitem{Ha03}
H.~I. Ha, J.~I. Oh, J. Moser, and M.~J. Naughton, Synth. Metals {\bf 137},
  1215  (2003).

\bibitem{Naughton05}
M. Naughton, private communication (unpublished).

\bibitem{Belin97}
S. Belin and K. Behnia, Phys. Rev. Lett. {\bf 79},  2125  (1997).

\bibitem{Shimahara00}
H. Shimahara, Phys. Rev. B {\bf 61},  R14936  (2000).

\bibitem{Fuseya05}
Y. Fuseya and Y. Suzumura, J. Phys. Soc. Jpn. {\bf 74},  1264  (2005).

\bibitem{Solyom79}
J. S\'olyom, Adv. Phys. {\bf 28},  201  (1979).

\bibitem{Emery83}
V.~J. Emery, J. de Phys. (Paris) Coll. {\bf 44},  C3  (1983).

\bibitem{Barisic70}
S. Bari\v{s}i\'c, J. Labb\'e, and J. Friedel, Phys. Rev. Lett. {\bf 25},  919
  (1970).

\bibitem{Su79}
W.~P. Su, J.~R. Schrieffer, and A.~J. Heeger, Phys. Rev. Lett. {\bf 42},  1698
  (1979).

\bibitem{Creuzet87}
F. Creuzet {\it et~al.}, Synth. Metals {\bf 19},  289  (1987).

\bibitem{Bourbonnais96}
C. Bourbonnais and B. Dumoulin, J. Phys. I {\bf 6},  1727  (1996).

\bibitem{Emery86}
V.~J. Emery, Synth. Metals {\bf 13},  21  (1986).

\bibitem{Kohn65}
W. Kohn and J.~M. Luttinger, Phys. Rev. Lett. {\bf 15},  524  (1965).

\bibitem{Beal86}
M.~T. B\'eal-Monod, C. Bourbonnais, and V.~J. Emery, Phys. Rev. B {\bf 34},
  7716  (1986).

\bibitem{Caron86}
L.~G. Caron and C. Bourbonnais, Physica {\bf 143B},  453  (1986).

\bibitem{Scalapino86}
D.~J. Scalapino, E. Loh, and J.~E. Hirsch, Phys. Rev. B {\bf 34},  8190
  (1986).

\bibitem{Scalapino87}
D.~J. Scalapino, E. Loh, and J.~E. Hirsch, Phys. Rev. B {\bf 35},  6694
  (1987).

\bibitem{Miyake86}
K. Miyake, S. Schmitt-Rink, and C.~M. Varma, Phys. Rev. B {\bf 34},  6554
  (1986).

\bibitem{Shimahara88}
H. Shimahara, J. Phys. Soc. Jpn. {\bf 58},  1735  (1988).

\bibitem{Kino99}
H. Kino and H. Kontani, J. Low Temp. Phys {\bf 117},  317  (1999).

\bibitem{Kuroki99}
K. Kuroki and H. Aoki, Phys. Rev. B {\bf 60},  3060  (1999).

\bibitem{Scalapino95}
D.~J. Scalapino, Phys. Rep. {\bf 250},  329  (1995).

\bibitem{Duprat01}
R. Duprat and C. Bourbonnais, Eur. Phys. J. B {\bf 21},  219  (2001).

\bibitem{Kuroki01}
K. Kuroki, R. Arita, and H. Aoki, Phys. Rev. B {\bf 63},  094509  (2001).

\bibitem{Onari04}
S. Onari, R. Arita, K. Kuroki, and H. Aoki, Phys. Rev. B {\bf 70},  94523
  (2004).

\bibitem{Tanaka04}
Y. Tanaka and K. Kuroki, Phys. Rev. B {\bf 70},  R060502  (2004).

\bibitem{Bourbonnais04}
C. Bourbonnais and R. Duprat, Bull. Am. Phys. Soc. {\bf 49},  1:179  (2004).

\bibitem{Nickel05a}
J.~C. Nickel, R. Duprat, C. Bourbonnais, and N. Dupuis, to appear in Phys. Rev.
  Lett., arXiv:cond-mat/0502614 (unpublished).

\bibitem{Nickel05b}
J.~C. Nickel, R. Duprat, C. Bourbonnais, and N. Dupuis, arXiv:cond-mat/0510744
  (unpublished).

\bibitem{Hasegawa86}
Y. Hasegawa and H. Fukuyama, J. Phys. Soc. Jpn. {\bf 55},  3978  (1986).

\bibitem{Montambaux88}
G. Montambaux, Phys. Rev. B {\bf 38},  4788  (1988).

\bibitem{Gorkov74}
L.~P. Gor'kov and I.~E. Dzyaloshinskii, Sov. Phys. JETP {\bf 40},  198  (1974).

\bibitem{Mihaly76}
L. Mih\'aly and J. S\'olyom, J. Low Temp. Phys {\bf 24},  579  (1976).

\bibitem{Lee77}
P.~A. Lee, T.~M. Rice, and R.~A. Klemm, Phys. Rev. B {\bf 15},  2984  (1977).

\bibitem{Menyhard77}
N. Menyh\'ard, Sol. State Comm. {\bf 21},  495  (1977).

\bibitem{Barisic85}
S. Bari\v{s}i\'c and A. Bjeli\v{s},  in {\em Theoretical Aspects of Band
  Structures and Electronic Properties of Pseudo-One-Dimensional Solids},
  edited by H. Kaminura (D. Reidel, Dordrecht, 1985), p.\ 49.

\bibitem{Pouget89}
J.~P. Pouget and R. Comes,  in {\em Charge Density Waves in Solids}, edited by
  L.~P. Gor'kov and G. Gr\"uner (Elsevier Science, Amsterdam, 1989), p.\ 85.

\bibitem{Shimahara00b}
H. Shimahara, J. Phys. Soc. Jpn. {\bf 69},  1966  (2000).

\end{thebibliography}

\end{document}